\documentclass[11pt,oneside,reqno]{amsart}
\usepackage{amsfonts,amssymb, amscd,amsmath,latexsym,amsbsy,bm}
\numberwithin{equation}{section}
\usepackage{mathrsfs}
\usepackage{dsfont}
 \usepackage[usenames,dvipsnames]{xcolor}
 \usepackage[titletoc]{appendix}

\usepackage[colorlinks,hyperindex, 
           bookmarks=true,bookmarksopen=true]{hyperref}
           \hypersetup    {
       linkcolor=black,
        urlcolor=MidnightBlue,
        citecolor=RedOrange,
   }

\usepackage{epic}
\usepackage{eepic}
\usepackage{graphicx}

\usepackage{pgf}
\usepackage{tikz}

\usetikzlibrary{shapes}
\usetikzlibrary{plotmarks}

\usepackage{cite}
\usepackage{enumerate}
\usepackage[curve]{xy}
\usepackage{stmaryrd}

\usepackage{mathtools}
\DeclarePairedDelimiter{\floor}{\lfloor}{\rfloor}

\newcommand{\ds}{\displaystyle}

\renewcommand{\author}[1]{\large\rm #1\\ \bigskip}
\renewcommand{\title}[1]{\bigskip\bigskip\Large\bf #1\bigskip\bigskip\\}

\newcommand{\Bigpsi}[3]{\phantom{\Psi}_2 \kern -.05em
\Psi_2\left(\genfrac{}{}{0pt}{}{#1}{#2}\biggl|#3\right)}

\newcommand{\bea}{\begin{eqnarray}}
\newcommand{\eea}{\end{eqnarray}}

\newcommand{\beq}{\begin{equation}}
\newcommand{\eeq}{\end{equation}}

\newcommand{\x}{{\boldsymbol{x}}}
\newcommand{\y}{{\boldsymbol{y}}}

\newcommand{\ii}{\mathsf{i}}

\newcommand{\ow}{\overline{\mathcal W}}
\newcommand{\w}{{\mathcal W}}
\newcommand{\s}{{\mathcal S}}

\newcommand{\cpar}{{\eta}}

\newcommand{\q}{{\mathsf q}}
\newcommand{\p}{{\mathsf p}}

\newcommand{\qt}{{\tilde{\mathsf q}}}
\newcommand{\qb}{{\overline{\mathsf q}}}

\newcommand{\iW}{\mathcal{W}}
\newcommand{\iS}{\mathcal{S}}

\renewcommand{\L}{{\mathscr L}}

\def\EXP{\textrm{{\large e}}}

\def\re{\mathop{\hbox{\rm Re}}\nolimits}
\def\im{\mathop{\hbox{\rm Im}}\nolimits}

\renewcommand{\textcolor}[1]{}

 {\begin{list}{}%
         {\setlength{\leftmargin}{#1}}%
         \item[]%
 }
 {\end{list}}

\usepackage{mathtext}
\usepackage{stackrel}

\usepackage[utf8]{inputenc}

\begin{document} 

$\;$\\ 
\vspace{2.7cm}

\begin{center}
{\Large \bf The star-triangle relation, lens partition function,  and hypergeometric sum/integrals} 
\end{center}

\vskip 1.3 cm

\centerline{\large {\bf Ilmar Gahramanov$^{a,b,c}$ and Andrew P. Kels$^d$}  }

{\small
\begin{center}
\textit{$^a$Max Planck Institute for Gravitational Physics (Albert Einstein Institute),\\ Am M\"{u}hlenberg 1, D-14476 Potsdam, Germany} \\
\vspace{2.5mm}
\textit{ $^b$Institute of Radiation Problems ANAS,\\ B.Vahabzade 9, AZ1143 Baku, Azerbaijan} \\
\vspace{2.5mm}
\textit{$^c$Department of Mathematics, Khazar University, \\ Mehseti St. 41, AZ1096, Baku, Azerbaijan} \\
\vspace{2.5mm}
\textit{$^d$Institute of Physics, University of Tokyo, \\ Komaba, Tokyo 153-8902, Japan} \\
\texttt{} \\
\vspace{.1mm}
\vspace{.1mm}
\end{center}
}

\vskip 0.5cm \centerline{\bf Abstract} \vskip 0.2cm \noindent  The aim of the present paper is to consider the hyperbolic limit of an elliptic hypergeometric sum/integral identity, and associated lattice model of statistical mechanics previously obtained by the second author.  The hyperbolic sum/integral identity obtained from this limit, has two important physical applications in the context of the so-called gauge/YBE correspondence.  For statistical mechanics, this identity is equivalent to a new solution of the star-triangle relation form of the Yang-Baxter equation, that directly generalises the Faddeev-Volkov models to the case of discrete and continuous spin variables.  On the gauge theory side, this identity represents the duality of lens ($S_b^3/\mathbb{Z}_r$) partition functions, for certain three-dimensional $\mathcal N = 2$ supersymmetric gauge theories.

\newpage

{\color{black}
\small
\vspace{-0.45cm}
{\parskip=0pt
\tableofcontents
}
\vspace{-0.25cm}
}
\normalsize

\section{Introduction}

In the last few years several interesting connections between integrable lattice models and supersymmetric gauge theories have been revealed \cite{Spiridonov:2010em,Yamazaki:2012cp,Yamazaki:2013nra, Yagi:2015lha,Yamazaki:2015voa,Gahramanov:2015cva,Nieri:2013vba,Costello:2013zra,Costello:2013sla,Maruyoshi:2016caf, Nekrasov:2009uh,Nekrasov:2009rc,Terashima:2012cx,Xie:2012mr}.  One of these connections known as the gauge/YBE correspondence, is a relationship between quiver gauge theories and integrable lattice models, where the integrability on the lattice side emerges as a manifestation of supersymmetric duality on the gauge theory side.  In the context of this correspondence, partition functions of supersymmetric quiver gauge theories are identified with partition functions of two-dimensional integrable lattice models of statistical mechanics.  This particular correspondence was initially noted by Spiridonov \cite{Spiridonov:2010em}, where he interpreted the star-triangle relation of Bazhanov, and Sergeev (BS) \cite{Bazhanov:2010kz}, in terms of dual superconformal indices for four-dimensional $\mathcal N=1$ quiver gauge theories \cite{Dolan:2008qi}, and was further developed by Yamazaki \cite{Yamazaki:2012cp,Yamazaki:2013nra}, who used four-dimensional $\mathcal{N}=1$ quiver gauge theory on the lens space \cite{Benini:2011nc}, to generalise the multi-spin integrable lattice model of BS \cite{Bazhanov:2011mz,Bazhanov:2013bh} satisfying the star-star relation.  

The most important advantage of the correspondence, is that in some cases new solutions of the Yang-Baxter equation are able to be systematically derived from calculations of supersymmetric gauge theory.  This aspect of the correspondence has proven to be a powerful tool in obtaining some quite general two-dimensional integrable lattice models \cite{Yamazaki:2013nra,Yamazaki:2015voa,Gahramanov:2015cva,Yagi:2015lha,Jafarzade}.  In connection with the correspondence, a new integrable lattice model satisfying the star-triangle relation form of the Yang-Baxter equation was recently discovered \cite{Kels:2015bda}, closely related to Yamazaki's solution of the star-star relation \cite{Yamazaki:2013nra}.   The model \cite{Kels:2015bda} generalises the BS master solution model \cite{Bazhanov:2010kz} to the case of discrete and continuous spin variables, and contains all known single-spin solutions of the star-triangle relation as special cases.  The star-triangle relation for this model was shown to result from a new type of elliptic hypergeometric sum/integral identity \cite{Kels:2015bda}, that may be considered a generalisation of Spiridonov's elliptic beta integral \cite{SpiridonovEBF} to the case of both complex and integer variables.  Just recently Spiridonov extended this elliptic sum/integral \cite{Kels:2015bda} to the multiple sum/integral case associated with the root system $C_n$, and described the corresponding analogue of the Gauss hypergeometric function \cite{rarified}.

The aim of the present work, is to consider details of the hyperbolic limit of the above elliptic hypergeometric sum/integral \cite{Kels:2015bda}, and associated lattice model satisfying star-triangle relation, in the context of the gauge/YBE correspondence.  This limit results in a hyperbolic analogue of the elliptic hypergeometric sum/integral, that notably has two different physical interpretations through this correspondence. First, it as a Yang-Baxter equation underpinning integrability of a new Ising type lattice model of statistical mechanics, generalising the Faddeev-Volkov models \cite{Bazhanov:2007mh,Bazhanov:2007vg,Spiridonov:2010em}, and second, it represents the equality of dual $S_b^3/\mathbb{Z}_r$ lens partition functions for three-dimensional $\mathcal N=2$ supersymmetric gauge theory \cite{Benini:2011nc}.

On the mathematical side, the hyperbolic sum/integral generalises the univariate hyperbolic beta integral \cite{STOKMAN2005119} to the case of both complex and integer variables.  Taking the limit from the elliptic identity \cite{Kels:2015bda} to the hyperbolic identity ends up being fairly straightforward, after utilising previous estimates in the hyperbolic limit given by Rains \cite{Rains2009}.  As in the latter cases the error introduced in taking the hyperbolic limit here is exponentially small, and the elliptic identity is found to converge exponentially quickly to the hyperbolic identity.  Both the new solution of the star-triangle relation, and corresponding duality of $S_b^3/\mathbb{Z}_r$ $\mathcal{N}=2$ partition functions considered in this paper, are obtained from the hyperbolic sum/integral identity after simple changes of variables.

The rest of the paper is organized as follows.  Section \ref{sec:statmechintro} provides a general overview of the type of two-dimensional lattice models of statistical mechanics considered throughout this paper.  In Section \ref{sec:newsolutions}, the explicit solutions of the star-triangle relation for models with discrete and continuous spin variables are given, along with some of their important properties.  Sections \ref{sec:gaugetheory}, and \ref{sec:ellipticgauge} show how such star-triangle relations may be obtained from supersymmetric gauge theory calculations, for the hyperbolic and elliptic cases respectively.

Some properties of the special functions used in this paper are summarised in Appendix \ref{app:functions}.  Appendix \ref{app:hyper} gives details of the hyperbolic limit of the elliptic hypergeometric sum/integral identity \cite{Kels:2015bda}, analogously to previous hyperbolic limits for elliptic hypergeometric integrals \cite{Rains2009}.  Appendix \ref{app:modr} describes another form of the elliptic and hyperbolic sum/integral identities, without dependence on the modulus r.  Particularly it is shown here that there is a certain freedom in the choice of normalisation of the lens elliptic gamma function, and as a consequence of this, the normalisation recently used in \cite{rarified}, that is different from the original normalisation \cite{Kels:2015bda}, results in exactly the same elliptic sum/integral identity.\footnote{This is contrary to remarks in \cite{rarified}, where it was incorrectly concluded that the different normalisation of the lens elliptic gamma function results in an identity that is different from \cite{Kels:2015bda}.}

\section{Two-dimensional exactly solved models of statistical mechanics}\label{sec:statmechintro}

The models of statistical mechanics considered in this paper, are Ising type models of interacting spins located at vertices of a two-dimensional lattice.  The models considered here are integrable, and satisfy a particular form of the Yang-Baxter equation known as the star-triangle relation.  This class of integrable models includes many important examples, such as the two-dimensional Ising \cite{Baxter:1982zz}, Fateev-Zamolodchikov \cite{Fateev:1982wi}, Kashiwara-Miwa \cite{Kashiwara:1986tu}, Chiral Potts models \cite{AuYang:1987zc,Baxter:1987eq}, and several others \cite{Zamolodchikov:1980mb,FV95,Bazhanov:2007mh,Bazhanov:2007vg,Bazhanov:2010kz,Kels:2013ola,Kels:2015bda,Baxter:1978xr,Bazhanov:2016ajm}.  Here a quite general overview of such integrable lattice models and their properties will be given, before moving on to the explicit new examples in Section \ref{sec:newsolutions}.

\subsection{Square lattice model}
\begin{figure}[htb]
\centering
\begin{tikzpicture}[scale=1]

\draw[-,thick] (-0.5,-0.5)--(3.5,3.5);
\draw[-,thick] (-0.5,3.5)--(3.5,-0.5);
\draw[-,thick] (-0.5,1.5)--(1.5,3.5)--(3.5,1.5)--(1.5,-0.5)--(-0.5,1.5);
\draw[-,thick] (-4.5,-0.5)--(-0.5,3.5);
\draw[-,thick] (-4.5,3.5)--(-0.5,-0.5);
\draw[-,thick] (-4.5,1.5)--(-2.5,3.5)--(-0.5,1.5)--(-2.5,-0.5)--(-4.5,1.5);
\foreach \x in {-4,...,3}{
\draw[->,dashed] (\x,-1) -- (\x,4);
\fill[white!] (\x,-1) circle (0.1pt)
node[below=0.05pt]{\color{black}\small $q$};}
\foreach \y in {0,...,3}{
\draw[->,dashed] (-5,\y) -- (4,\y);
\fill[white!] (-5,\y) circle (0.1pt)
node[left=0.05pt]{\color{black}\small $p$};}
\foreach \y in {-0.5,1.5,3.5}{
\foreach \x in {-4.5,-2.5,...,3.5}{
\filldraw[fill=black,draw=black] (\x,\y) circle (2.2pt);}}
\foreach \y in {0.5,2.5}{
\foreach \x in {-3.5,-1.5,...,2.5}{
\filldraw[fill=black,draw=black] (\x,\y) circle (2.2pt);}}

\end{tikzpicture}
\caption{A square lattice $L$ (solid lines) and its associated directed rapidity graph $\L$ (dashed lines).}
\label{squarelattice}
\end{figure}
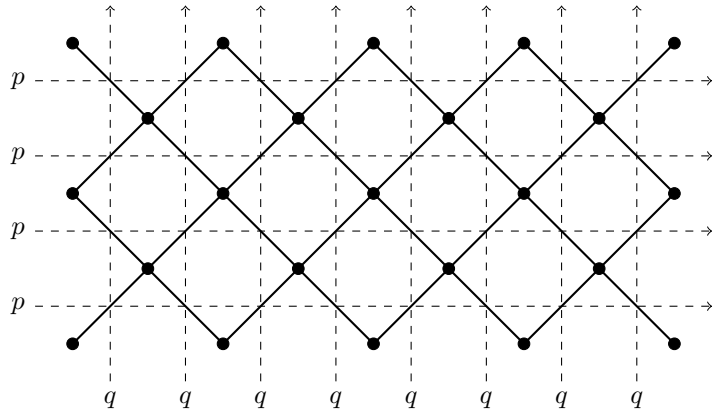

Introduce the square lattice $L$, that contains $N$ vertices, as is depicted graphically in Figure \ref{squarelattice}.

Each vertex $j$ of the lattice $L$, is assigned a spin variable, denoted $\sigma_j$, which takes some set of values.  Here the spins will be of the form
\beq
\label{spinintrodef}
\sigma_j=(x_j,m_j)\,,\quad j=1,2,\ldots,N,
\eeq
where the spin component $x_j$ takes values in some subset of $\mathbb{R}$, and the spin component $m_j$ takes values in some subset of $\mathbb{Z}$.

\begin{figure}[hbt]
\centering
\begin{tikzpicture}[scale=2.6]

\draw[-,very thick] (-0.5,2)--(0.5,2);
\draw[->,dashed] (0.4,1.6)--(-0.4,2.4);
\fill[white!] (0.4,1.6) circle (0.01pt)
node[below=0.5pt]{\color{black}\small $q$};
\draw[->,dashed] (-0.4,1.6)--(0.4,2.4);
\fill[white!] (-0.4,1.6) circle (0.01pt)
node[below=0.5pt]{\color{black}\small $p$};
\filldraw[fill=black,draw=black] (-0.5,2) circle (0.9pt)
node[left=3pt]{\color{black} $\sigma_i$};
\filldraw[fill=black,draw=black] (0.5,2) circle (0.9pt)
node[right=3pt]{\color{black} $\sigma_j$};

\fill (0,1.3) circle(0.01pt)
node[below=0.05pt]{\color{black} $\w_{pq}(\sigma_i,\sigma_j)$};

\begin{scope}[xshift=60pt,yshift=57pt]
\draw[-,very thick] (0,-0.5)--(0,0.5);
\draw[->,dashed] (-0.4,-0.4)--(0.4,0.4);
\fill[white!] (-0.4,-0.4) circle (0.01pt)
node[below=0.5pt]{\color{black}\small $p$};
\draw[->,dashed] (0.4,-0.4)--(-0.4,0.4);
\fill[white!] (0.4,-0.4) circle (0.01pt)
node[below=0.5pt]{\color{black}\small $q$};
\filldraw[fill=black,draw=black] (0,-0.5) circle (0.9pt)
node[below=3pt]{\color{black} $\sigma_i$};
\filldraw[fill=black,draw=black] (0,0.5) circle (0.9pt)
node[above=3pt]{\color{black} $\sigma_j$};

\fill (0,-0.7) circle(0.01pt)
node[below=0.05pt]{\color{black} $\ow_{pq}(\sigma_i,\sigma_j)$};
\end{scope}
\end{tikzpicture}
\caption{Edges of the first (left) and second (right) types in the square lattice of Figure \ref{squarelattice}, and associated Boltzmann weights $\w_{pq}(\sigma_i,\sigma_j)$, and $\ow_{pq}(\sigma_i,\sigma_j)$.}
\label{2boltzmannweights}
\end{figure}
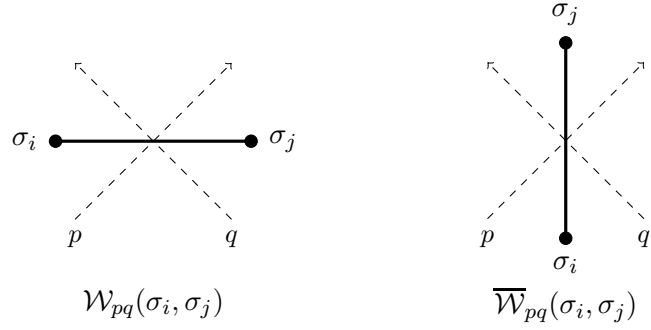

The directed rapidity graph $\L$, is represented in Figure \ref{squarelattice} by directed dashed lines crossing the edges of $L$ at 45 degree angles.  Two real valued rapidity variables, labelled $p$, and $q$, are respectively associated to horizontally and vertically directed rapidity lines.  The crossing of rapidity lines distinguishes two types of edges of the square lattice $L$, that are depicted graphically in Figure \ref{2boltzmannweights}.

The lattice model involves nearest neighbour interactions, where two spins $\sigma_i,\sigma_j$, interact only if they are at two vertices $i$, $j$, connected by an edge $(ij)$ of the lattice $L$.  The interactions are characterised by the Boltzmann weights $\w_{pq}(\sigma_i,\sigma_j)$, and $\ow_{pq}(\sigma_i,\sigma_j)$, associated to the two types of respective edges, where $\sigma_i$ and $\sigma_j$ are the spins located at opposite ends of an edge, as shown in Figure \ref{2boltzmannweights}.  Each of these Boltzmann weights generally depend on the value of the two spin variables, and the two rapidity variables, associated to an edge of $L$.

In all cases considered here, the two Boltzmann weights $\w_{pq}$, $\ow_{pq}$, depend only on the difference of rapidity variables $p-q$ (the majority of lattice models of statistical mechanics satisfy this property, the most notable exception being the Chiral Potts model \cite{Baxter:1987eq}).  Consequently the Boltzmann weights will be written in terms of the spectral variable $\alpha=p-q$, as 
\begin{equation}
\w_\alpha(\sigma_i,\sigma_j):=\w_{pq}(\sigma_i,\sigma_j)\,, \quad \text{and} \quad\, \ow_\alpha(\sigma_i,\sigma_j):=\ow_{pq}(\sigma_i,\sigma_j) \,.
\end{equation}
The two Boltzmann weights are also related by the crossing symmetry 
\begin{equation}
\ow_\alpha(\sigma_i,\sigma_j)=\w_{\eta-\alpha}(\sigma_i,\sigma_j) \,,
\end{equation}
where $\eta>0$ is a real valued, model dependent ``crossing parameter''.  Thus all two-spin interactions in the lattice model may be described in terms of the single Boltzmann weight $\w_\alpha(\sigma_i,\sigma_j)$. 

The Boltzmann weights considered here are spin reflection symmetric, such that $\w_\alpha(\sigma_i,\sigma_j)=\w_\alpha(\sigma_j,\sigma_i)$.  Importantly, all lattice models considered here may be interpreted as ``physical'', such that all interactions are described by positive, real valued Boltzmann weights $\w_\alpha(\sigma_i,\sigma_j)$, that represent real valued interaction energies.

The model also depends on the single-spin Boltzmann weight $\s(\sigma_j)$, associated to each vertex $j$ of the lattice.  This Boltzmann weight depends only on the value of the spin $\sigma_j$, and is independent of any rapidity variables.

\subsection{Partition function and star-triangle relation}
The {\it partition function} for the above lattice model is given by the expression
\begin{align}
\label{z-main}
{\mathcal Z}=\sum\int
\prod_{(ij)}\w_{\alpha}(\sigma_i,\sigma_j)\
\prod_{(kl)}\w_{\cpar-\alpha}(\sigma_k,\sigma_l)\ \prod_{n}
\s(\sigma_n)\,d x_n\,.
\end{align}
In this expression, the first product is taken over all edges $(ij)$ of the first type in Figure \ref{2boltzmannweights}, the second over all edges $(kl)$ of the second type in Figure \ref{2boltzmannweights}, and the third product over all interior vertices $n$ of the lattice $L$.  The integral and sum are taken over all possible values of interior spins $\sigma_n=(x_n,m_n)$ in the lattice (these values depend on the actual definition of spins \eqref{spinintrodef} for the particular model), and boundary spins are assigned fixed values.   

\begin{figure}[tbh]
\centering
\begin{tikzpicture}[scale=2]

\draw[-,very thick] (-2,0)--(-2,1);
\draw[-,very thick] (-2,0)--(-2.87,-0.5);
\draw[-,very thick] (-2,0)--(-1.13,-0.5);
\draw[->,black,dashed] (-2.9,-0.25)--(-1.1,-0.25);
\fill[white!] (-2.9,-0.25) circle (0.5pt)
node[left=1.5pt]{\color{black}\small $p$};
\draw[->,black,dashed] (-2.7,-0.71)--(-1.7,1.02);
\fill[white!] (-2.7,-0.71) circle (0.5pt)
node[below=1.5pt]{\color{black}\small $q$};
\draw[->,black!30!black,dashed] (-1.3,-0.71)--(-2.3,1.02);
\fill[white!] (-1.3,-0.74)circle (0.5pt)
node[below=1.5pt]{\color{black}\small $r$};
\fill (-2,0) circle (1.2pt)
node[below=2.5pt]{\color{black} $\sigma_a$};
\filldraw[fill=black,draw=black] (-2,1) circle (1.2pt)
node[above=1.5pt] {\color{black} $\sigma_i$};
\filldraw[fill=black,draw=black] (-2.87,-0.5) circle (1.2pt)
node[left=1.5pt] {\color{black} $\sigma_k$};
\filldraw[fill=black,draw=black] (-1.13,-0.5) circle (1.2pt)
node[right=1.5pt] {\color{black} $\sigma_j$};

\fill[white!] (0.05,0.3) circle (0.01pt)
node[left=0.05pt] {\color{black}$=$};

\draw[-,very thick] (2,1)--(1.13,-0.5);
\draw[-,very thick] (1.13,-0.5)--(2.87,-0.5);
\draw[-,very thick] (2.87,-0.5)--(2,1);
\draw[->,black,dashed] (1.1,0.25)--(2.85,0.25);
\fill[white!] (1.1,0.25) circle (0.5pt)
node[left=1.5pt]{\color{black}\small $p $};
\draw[->,black,dashed] (1.75,-0.93)--(2.68,0.67);
\fill[white!] (1.75,-0.93) circle (0.5pt)
node[below=1.5pt]{\color{black}\small $q$};
\draw[->,black!30!black,dashed] (2.25,-0.93)--(1.32,0.67);
\fill[white!] (2.25,-0.96) circle (0.5pt)
node[below=1.5pt]{\color{black}\small $r$};
\filldraw[fill=black,draw=black] (2,1) circle (1.2pt)
node[above=1.5pt]{\color{black} $\sigma_i$};
\filldraw[fill=black,draw=black] (1.13,-0.5) circle (1.2pt)
node[left=1.5pt]{\color{black} $\sigma_k$};
\filldraw[fill=black,draw=black] (2.87,-0.5) circle (1.2pt)
node[right=1.5pt]{\color{black} $\sigma_j$};

\end{tikzpicture}
\caption{The star-triangle relation \eqref{msstr}.}
\label{STR-figure}
\end{figure}
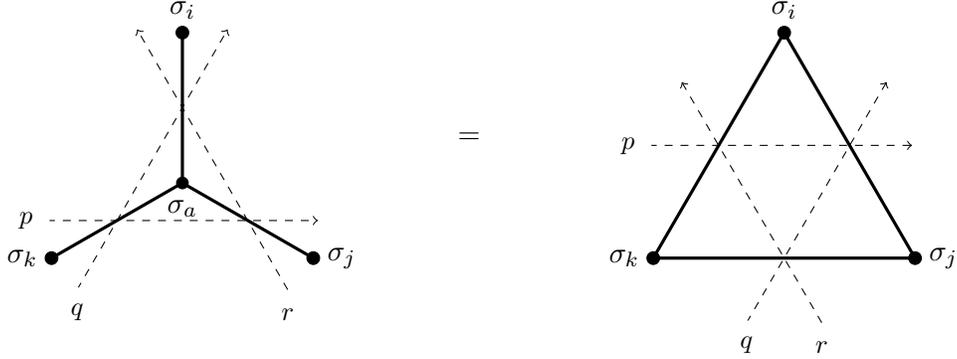

The goal of statistical mechanics \cite{Baxter:1982zz} is to evaluate \eqref{z-main} in the thermodynamic limit, when $N\rightarrow\infty$.  An exact evaluation is possible if the Boltzmann weights satisfy the Yang-Baxter equation \cite{Baxter:1972hz}, which for models considered here takes the form of the following {\it star-triangle relation}
\begin{align}
\label{msstr}
\begin{array}{l}
\ds\sum_{m_a}\int\!
 d x_a\,\iS(\sigma)\iW_{\eta-\alpha_i}(\sigma_i,\sigma_a)\iW_{\eta-\alpha_j}(\sigma_j,\sigma_a)\iW_{\eta-\alpha_k}(\sigma_k,\sigma_a)\\[.3cm]
\phantom{MMMMMMMMM}\ds=
{\mathcal
  R}(\alpha_i,\alpha_j,\alpha_k)\,\iW_{\alpha_i}(\sigma_j,\sigma_k)\iW_{\alpha_j}(\sigma_i,\sigma_k)\iW_{\alpha_k}(\sigma_j,\sigma_i)\,,
\end{array}
\end{align}
that is depicted graphically in Figure \ref{STR-figure}.

Here the three spectral parameters satisfy a constraint $\alpha_i+\alpha_j+\alpha_k=\eta$, and the factor ${\mathcal   R}(\alpha_i,\alpha_j,\alpha_k)$ is independent of any spin variables.  The integral and sum are evaluated over all possible values of the interior spin $\sigma_a=(x_a,m_a)$, while the boundary spins $\sigma_i,\sigma_j,\sigma_k$, are kept fixed.

A second star-triangle relation is also required, that is obtained by reversing the orientation of each rapidity line appearing in Figure \ref{STR-figure}.  However for models considered here satisfying reflection symmetry (such that $\w_\alpha(\sigma_i,\sigma_j)=\w_\alpha(\sigma_j,\sigma_i)$), the second expression is equivalent to \eqref{msstr}.  The key point is that the star-triangle relation \eqref{msstr} allows for an exact evaluation of \eqref{z-main} in the thermodynamic limit, with the use of the commuting transfer matrices technique pioneered by Baxter \cite{Baxter:1972hz,Baxter:1982zz}.

\subsection{Boltzmann weight normalisation and inversion relations}
For all models considered here, a normalisation of the Boltzmann weights is chosen such that $\mathcal{R}(\alpha_i,\alpha_j,\alpha_k)=1$ \cite{Bazhanov:2016ajm,Bax02rip}.  For this particular normalisation, the Boltzmann weights of the model may be shown to satisfy the following boundary conditions (for other normalisations, some extra factors will appear on the right hand sides of \eqref{msbc})
\begin{align}
\label{msbc}
\begin{array}{rcl}
\ds\left.\w_\alpha(\sigma_i,\sigma_j)\right|_{\alpha=0}&=&\ds1,\\[0.3cm]
\ds\left.\w_{\eta-\alpha}(\sigma_i,\sigma_j)\right|_{\alpha\rightarrow0}&=&\ds\frac{1}{2\s(\sigma_i)}(\delta(x_i\!+\!x_j)\,\delta_{m_i,-m_j}+\delta(x_i\!-\!x_j)\,\delta_{m_i,m_j})\,,
\end{array}
\end{align}
for all values of the spins $\sigma_i,\sigma_j$, where $\s(\sigma_i)\neq0$, and $\delta(x)$, and $\delta_{m,n}$, are respectively Dirac and Kronecker delta functions.  The exact form of the second boundary condition differs slightly depending on the symmetries satisfied by the Boltzmann weights, and explicit expressions will be given for the two cases considered in Section \ref{sec:newsolutions}.

The boundary conditions \eqref{msbc}, and star-triangle relation \eqref{msstr}, imply the following {\it inversion relations}
\begin{align}
\label{msinv}
\begin{array}{rcl}
\ds\w_\alpha(\sigma_i,\sigma_j)\w_{-\alpha}(\sigma_i,\sigma_j)&=&1\,, \\[0.3cm]
\ds\sum_{m_0}\!\int\!\! dx_0\,\s(\sigma_0)\w_{\eta-\alpha}(\sigma_i,\sigma_0)\w_{\eta+\alpha}(\sigma_0,\sigma_j)&=&\ds\frac{1}{\s(\sigma_i)}(\delta(x_i\!+\!x_j)\,\delta_{m_i,-m_j}+\delta(x_i\!-\!x_j)\,\delta_{m_i,m_j})\,.
\end{array}
\end{align}
The exact form of the inversion relations again depend on the model being considered.  The above relations, \eqref{msstr} and \eqref{msinv}, may be used to show that in the thermodynamic limit when $N\to\infty$, the bulk free energy of the model vanishes
\begin{align}
 \lim_{N\to \infty} N^{-1} \log{\mathcal  Z} = 0\,.\label{fzero}
\end{align}
A derivation of this result requires some extensions \cite{Bazhanov:2016ajm} of the standard inversion relation method \cite{Stroganov:1979et,Zamolodchikov:1979ba,Baxter:1982xp}.  Here the boundary spins are assumed to be kept finite in the limit $N\to\infty$, and there is an analyticity assumption for the free energy of the model in the physical regime.  Note that the result \eqref{fzero} is purely a consequence of the special choice of normalisation for the Boltzmann weights \cite{Bazhanov:2007mh,Bazhanov:2007vg,Bazhanov:2010kz}, {\it i.e.} the free energy is contained in the normalisation.  Another advantage \cite{Bazhanov:2016ajm} of using the normalisation \eqref{msbc}, is strict invariance of the partition function \eqref{z-main} under deformations of the rapidity lattice $\L$, associated with Z-invariance \cite{Baxter:1978xr}, however this property will not be required in the following.

\section{Solutions of the star-triangle relation with continuous and discrete spins}\label{sec:newsolutions}

\subsection{The elliptic case}\label{sec:ellipticmodel}

The most general known solution of the star-triangle relation for Ising type models with scalar valued spin components, was recently discovered by the second author \cite{Kels:2015bda}.  The corresponding lattice model is a generalisation of the Bazhanov and Sergeev (BS) master solution of the star-triangle relation \cite{Bazhanov:2010kz}, to the case of discrete and continuous spin variables.  The star-triangle relation of the model arises as a particular case of a new type of elliptic hypergeometric sum/integral identity, that generalises Spiridonov's elliptic beta integral \cite{SpiridonovEBF} to the case of both integer and complex variables.

A review of the lattice model and corresponding star-triangle relation \cite{Kels:2015bda} is given below.  On the gauge theory side, the Boltzmann weights for this model originally appeared in Yamazaki's elliptic solution of the {\it star-star relation} \cite{Yamazaki:2013nra}, that corresponds to dualities of four-dimensional $\mathcal{N}=1$ supersymmetric quiver gauge theory \cite{Benini:2011nc}.  The star-triangle relation \cite{Kels:2015bda} implies the particular duality \cite{Yamazaki:2013nra} for the case of $SU(2)$ gauge group and corresponding star-star relation.  However there is no known star-triangle relation for the general $SU(N)$ case and corresponding star-star relations involving vector valued spins.  The interpretation of this model \cite{Kels:2015bda} in terms of four-dimensional $\mathcal{N}=1$ quiver gauge theory will be discussed further in Section \ref{sec:ellipticgauge}.

The following spins
\beq
\label{spindef}
\sigma_j=(x_j,m_j),\quad 0\leq x_j<\pi,\quad m_j=0,1,\ldots,\floor{r/2}\,,
\eeq
are assigned to each vertex $j$ of the lattice, where $r\in\{1,2,\ldots,\}$ is a positive integer parameter, and $\floor{~}$ is the floor function.  The model depends on two elliptic nomes $\p$, and $\q$, which are defined as
\beq
\label{nomes}
\p=\EXP^{\pi\ii\sigma},\;\q=\EXP^{\pi\ii\tau},\quad\im\sigma,\;\im\tau >0\,.
\eeq
The elliptic nomes act as temperature like parameters, and taking an elliptic nome to the unit circle corresponds to a ground state of the model, that is evaluated on the solution of a {\it classical} discrete integrable equation \cite{Bazhanov:2010kz,Bazhanov:2011mz,Bazhanov:2016ajm}.  The crossing parameter is defined in terms of $\sigma$, and $\tau$ as
\beq
\eta=-\pi\ii(\sigma+\tau)/2\,.
\eeq
The crossing parameter $\eta$ is required to be real and positive, which corresponds to a physical regime of the model.

The elliptic gamma function \cite{Ruijsenaars:1997:FOA} is defined here as 
\beq
\label{egf}
\Phi(z;\p,\q)=\prod_{j,k=0}^\infty\frac{1-\EXP^{2\ii z}\,\p^{2j+1}\,\q^{2k+1}}{1-\EXP^{-2\ii z}\,\p^{2j+1}\,\q^{2k+1}}\,.
\eeq
Notably, the expression for the elliptic gamma function \eqref{egf} appears implicitly \cite{Spiridonov-essays} in Baxter's solution of the eight-vertex model \cite{Baxter:1972hz}, while studies of such generalised gamma functions were initiated over 100 years ago with the work of Barnes \cite{Barnes:1901}.

The ``lens'' elliptic gamma function is defined as the following product of two elliptic gamma functions
\beq
\label{legf}
\begin{array}{rcl}
\ds\Phi_{r,m}(z)&=&\ds\Phi(z+(r/2-\llbracket m\rrbracket _r)\,\pi\sigma;\p\,\q,\,\p^r)\,\Phi(z-(r/2-\llbracket m\rrbracket _r)\,\pi\tau;\p\,\q,\,\q^r) \\[0.3cm]
&=&\ds\prod_{j,k=0}^\infty\!\frac{1-\EXP^{2\ii z}\,\p^{-2\llbracket m\rrbracket _r}\,(\p\q)^{2j+1}\,(\p^r)^{2k+2}}{1-\EXP^{-2\ii z}\,\p^{2\llbracket m\rrbracket _r}\,(\p\q)^{2j+1}\,(\p^r)^{2k}}\frac{1-\EXP^{2\ii z}\,\q^{2\llbracket m\rrbracket _r}\,(\p\q)^{2j+1}\,(\q^r)^{2k}}{1-\EXP^{-2\ii z}\,\q^{-2\llbracket m\rrbracket _r}\,(\p\q)^{2j+1}\,(\q^r)^{2k+2}}\,,
\end{array}
\eeq
where $\llbracket  m\rrbracket _r\in\{0,1,\ldots ,r-1\}$ denotes $m \mbox{ modulus } r$.  This function originated in studies of the superconformal indices of four-dimensional $\mathcal{N}=1$ gauge theories involving the lens space \cite{Yamazaki:2013nra,Yamazaki:2012cp,Yamazaki:2013fva,Benini:2011nc}. If $r=1$, then $\llbracket m\rrbracket_1=0$ for any $m$, and the lens elliptic gamma function reduces to the usual elliptic gamma-function \eqref{egf}
\beq
\Phi_{1,0}(z)=\Phi(z;\p,\q)\,.
\eeq
The lens elliptic gamma function \eqref{legf}, satisfies the following periodicity and inversion relations
\beq
\Phi_{r,m}(z)=\Phi_{r,m}(z+\pi),\qquad\frac{1}{\Phi_{r,m}(z)}=\Phi_{r,-m}(-z)\,.
\eeq
The two-spin Boltzmann weight of the lattice model, is defined in terms of the lens elliptic gamma function \eqref{legf} as
\beq
\label{ebw}
\w_\alpha(\sigma_i,\sigma_j)=\ds\frac{\EXP^{-\frac{2\alpha}{r}(\,\llbracket m_i-m_j\rrbracket _\pm+\llbracket m_i+m_j\rrbracket _\pm)}}{\kappa^e(\alpha)}\frac{\Phi_{r,m_i-m_j}(x_i-x_j+\ii\alpha)\,\Phi_{r,m_i+m_j}(x_i+x_j+\ii\alpha)}{\Phi_{r,m_i-m_j}(x_i-x_j-\ii\alpha)\,\Phi_{r,m_i+m_j}(x_i+x_j-\ii\alpha)}\,,
\eeq
where $\llbracket m\rrbracket_\pm:=\llbracket m\rrbracket_r\llbracket -m\rrbracket_r$.  In the physical regime this represents the energy of the interaction between two spins $\sigma_i,\sigma_j$, at vertices connected by an edge of the lattice.  The Boltzmann weight \eqref{ebw} also depends on the spectral parameter $\alpha$, which is restricted to values $0\leq\alpha<\eta$.  For $\p=\q^*$, the Boltzmann weights \eqref{ebw} are positive and real valued, corresponding to a physical regime of the model.

The normalisation factor $\kappa^e(\alpha)$ in \eqref{ebw} is defined as (the superscript $e$ is introduced here to distinguish the normalisation $\kappa^e(\alpha)$, from the normalisation $\kappa^h(\alpha)$ introduced in the next subsection)
\beq
\label{ebwnorm}
\kappa^e(\alpha)=\exp\left\{\sum_{n\neq0}\frac{\EXP^{4\alpha n}((\p\q)^{rn}-(\p\q)^{-rn})}{n((\p\q)^{2n}-(\p\q)^{-2n})(\p^{rn}-\p^{-rn})(\q^{rn}-\q^{-rn})}\right\}.
\eeq
This function satisfies the required pair of functional equations
\beq
\label{functrels}
\frac{\kappa^e(\eta-\alpha)}{\kappa^e(\alpha)}=\Phi_{r,0}(\ii(\eta-2\alpha)),\quad\kappa^e(\alpha)\kappa^e(-\alpha)=1\,,
\eeq
so that the factor $\mathcal{R}(\alpha_i,\alpha_j,\alpha_k)=1$ in \eqref{msstr}.  The function $\kappa^e(\alpha)$ thus represents the {\it partition function per edge} of the square lattice model, obtained through the inversion relation method \cite{Stroganov:1979et,Zamolodchikov:1979ba,Baxter:1982xp,Bazhanov:2016ajm}, and for $r=1$ reduces to the partition function per edge function appearing for the BS master solution \cite{Bazhanov:2010kz}.

The one-spin Boltzmann weight of the model is defined as
\beq
\label{ebws}
\begin{array}{rcl}
\ds\s(\sigma_i)&=&\ds\frac{\varepsilon_i}{\pi}\,(\p^{2r};\p^{2r})_\infty(\q^{2r};\q^{2r})_\infty\,\EXP^{2\eta\llbracket 2m_i\rrbracket _\pm/r}\,\Phi_{r,-2m_i}(-2x_i-\ii\eta)\,\Phi_{r,2m_i}(2x_i-\ii\eta)\,, \\[0.5cm]
&=&\ds\frac{\varepsilon_i}{\pi}\,\EXP^{2\eta\llbracket 2m_i\rrbracket _\pm/r}\,{\vartheta}_4(2x_i+(r/2-\llbracket 2m_i\rrbracket _r)\pi\sigma\,|\,\p^r)\,{\vartheta}_4(2x_i-(r/2-\llbracket 2m_i\rrbracket _r)\pi\tau\,|\,\q^r)\,,
\end{array}
\eeq
where
\beq
\label{epsdef}
\varepsilon_i=\left\{\begin{array}{ll}\frac{1}{2}&\ds\quad m_i=0 \mbox{ or } \llbracket r-m_i\rrbracket _r\,, \\[0.3cm] 1&\quad\mbox{otherwise}\,,\end{array}\right.
\eeq
$\vartheta_4(z\,|\,\p)$ is a Jacobi theta function
\beq
\vartheta_4(z\,|\,\p)=(\p^2;\p^2)_\infty\prod_{n=1}^\infty\left(1-\EXP^{2\ii z}\p^{2n-1}\right)\left(1-\EXP^{-2\ii z}\p^{2n-1}\right),
\eeq
and
\beq
(x;\q)_\infty=\prod_{j=0}^\infty\,(1-x\,\q^j)\,,
\eeq
is the $\q$-Pochhammer symbol.

The Boltzmann weights \eqref{ebw} are spin reflection symmetric, such that
\beq
\label{spinrefl}
\w_\alpha(\sigma_i,\sigma_j)=\w_\alpha(\sigma_j,\sigma_i)\,.
\eeq
The Boltzmann weights are $\pi$-periodic in the continuous spin variable, and they are invariant under the spin transformation $x_i\rightarrow-x_i$, $m_i\rightarrow r-m_i$,
The discrete spins are restricted to values $0,1,\ldots,\floor{r/2}$, and the $\varepsilon_i$ factor is introduced in \eqref{epsdef} in order to account for this.

The Boltzmann weights \eqref{ebw} satisfy the following boundary conditions analogous to \eqref{msbc}
\beq
\begin{array}{rcl}
\ds\left.\w_\alpha(\sigma_i,\sigma_j)\right|_{\alpha=0}&=&\ds1,\\[0.3cm]
\ds\left.\w_{\eta-\alpha}(\sigma_i,\sigma_j)\right|_{\alpha\rightarrow0}&=&\ds\frac{\varepsilon_{i}}{\s(\sigma_i)} \left(\delta(\sin(x_i\!+\!x_j))\,\delta_{\llbracket m_i+m_j\rrbracket_r,0}+\delta(\sin(x_i\!-\!x_j))\,\delta_{\llbracket m_i-m_j\rrbracket_r,0}\right),
\end{array}
\eeq
where $\s(\sigma_i)\neq0$. 

The Boltzmann weights \eqref{ebw}, and \eqref{ebws}, satisfy the following star-triangle relation \cite{Kels:2015bda}
\beq
\label{str}
\begin{array}{r}
\ds\sum_{m_0=0}^{\floor{r/2}}\,\int^\pi_0\! dx_0\,\iS(\sigma_0)\,\iW_{\eta-\alpha_i}(\sigma_i,\sigma_0)\,\iW_{\eta-\alpha_j}(\sigma_j,\sigma_0)\,\iW_{\eta-\alpha_k}(\sigma_k,\sigma_0)\phantom{\,,} \\[0.4cm]
\ds=\iW_{\alpha_i}(\sigma_j,\sigma_k)\,\iW_{\alpha_j}(\sigma_i,\sigma_k)\,\iW_{\alpha_k}(\sigma_j,\sigma_i)\,,
\end{array}
\eeq
with the spectral parameters satisfying $\eta=\alpha_i+\alpha_j+\alpha_k$.  This star-triangle relation arises as particular case of the elliptic hypergeometric sum/integral \eqref{eident} \cite{Kels:2015bda}, which for $r=1$ is equivalent to Spiridonov's elliptic beta integral identity \cite{SpiridonovEBF}.

The expression for the partition function of the lattice model is given by \eqref{z-main}, with the Boltzmann weights defined in \eqref{ebw}, and \eqref{ebws}, and the integral and sum taken over all values of spins defined in \eqref{spindef}.  The star-triangle relation \eqref{str} corresponds to Seiberg duality of indices for four-dimensional $\mathcal{N}=1$ quiver gauge theories on the lens space, which will be discussed in more detail in Section \ref{sec:ellipticgauge}.  For the specific case of $r=1$, the star-triangle relation \eqref{str} is equivalent to the BS master solution of the star-triangle relation \cite{Bazhanov:2010kz}, while on the gauge theory side, \eqref{str} represents a duality in terms of four-dimensional $\mathcal{N}=1$ superconformal indices given by Dolan and Osborn \cite{Dolan:2008qi}.

\subsection{The hyperbolic case}\label{sec:hyperbolicmodel}

\subsubsection{Hyperbolic limit}
The hyperbolic limit of the star-triangle relation \eqref{str} when $r=1$ is well known \cite{STOKMAN2005119,VanDiejen2005,Rains2009,BultThesis}, and essentially involves directly replacing elliptic gamma functions \eqref{egf} with hyperbolic gamma functions.  The $r>1$ case turns out to be quite analogous, and as is to be be expected, involves introducing a generalisation of the hyperbolic gamma function (equivalently the non-compact quantum dilogarithm) obtained as the limit of the lens elliptic gamma function \eqref{legf}.  The generalisation of the hyperbolic gamma function, and some of its properties are summarised in Appendix \ref{app:functions}, while more details of the hyperbolic limit in terms of the elliptic hypergeometric sum/integral identity \cite{Kels:2015bda} corresponding to \eqref{str} may be found in Appendix \ref{app:hyper}.

Introduce the complex parameters $\omega_1,\omega_2$, where $\re(\omega_1),\re(\omega_2)>0$, and consider the following hyperbolic limit of the elliptic nomes \eqref{nomes}
\beq
\label{hyplim1}
\p=\EXP^{-\omega_1\epsilon}\,,\;\q=\EXP^{-\omega_2\epsilon},\quad\epsilon\rightarrow 0^+\,.
\eeq
This limit of the lens elliptic gamma function \eqref{legf} gives\footnote{This is an analogue of Proposition III.12 of Ruijsenaars \cite{Ruijsenaars:1997:FOA}.}
\beq \label{hyplim2}
\lim_{\epsilon\rightarrow0}\,\EXP^{\ii\pi^2 z/(6r\omega_1\omega_2\epsilon)}\,\Phi_{r,m}(z\epsilon)=\varphi_{r,m}(z)\,,
\eeq
where $m=0,1,\ldots,r-1$.  The function $\varphi_{r,m}(z)$ is defined for
\beq
-\re(\eta)-\min(\re(\omega_1)(r-\llbracket m\rrbracket),\re(\omega_2)\llbracket m\rrbracket)<\im(z)<\re(\eta)+\min(\re(\omega_1)\llbracket m\rrbracket,\re(\omega_2)(r-\llbracket m\rrbracket))\,,
\eeq
as
\beq
\label{hgammar}
\begin{array}{l}
\ds\varphi_{r,m}(z)=\exp\left\{\int_0^\infty dx\left(\frac{\ii z}{\omega_1\omega_2 rx^2}-\frac{\sinh(2\ii zx-\omega_1(r-2\llbracket m\rrbracket)x)}{2x\sinh(\omega_1rx)\sinh(2\eta x)}-\frac{\sinh(2\ii zx+\omega_2(r-2\llbracket m\rrbracket)x)}{2x\sinh(\omega_2rx)\sinh(2\eta x)}\right)\!\right\},
\end{array}
\eeq
where $\eta=(\omega_1+\omega_2)/2$.  From this definition it is straightforward to see that 
\beq
\varphi_{r,m}(z)\varphi_{r,-m}(-z)=1\,,
\eeq
and for $r=1$, $\varphi_{r,m}(z)$ is equivalent to the so-called hyperbolic gamma function, $\varphi_{1,0}(z;\omega_1,\omega_2)$, where
\beq
\varphi_{1,0}(z;\omega_1,\omega_2)=\exp\left\{\int^\infty_0dx\left\{\frac{\ii z}{\omega_1\omega_2x^2}-\frac{\sinh(2\ii zx)}{2x\sinh(\omega_1x)\sinh(\omega_2x)}\right)\right\}.
\eeq
Note that the usual convention for the hyperbolic gamma function is equivalent to $\varphi_{1,0}(-z;\omega_1,\omega_2)$ \cite{Ruijsenaars:1997:FOA}.  The function $\varphi_{r,m}(z)$, may also be written as the following product of two of the above hyperbolic gamma functions
\beq
\varphi_{r,m}(z)=\varphi_{1,0}(z+\ii\omega_1(r-2\llbracket m\rrbracket)/2;\omega_1,2\eta)\,\varphi_{1,0}(z-\ii\omega_2(r-2\llbracket m\rrbracket)/2;\omega_2,2\eta)\,.
\eeq
Further properties of $\varphi_{r,m}(z)$ are summarised in Appendix \ref{app:functions}.

In the limit \eqref{hyplim1}, the normalisation function $\kappa^e(\alpha)$ \eqref{ebwnorm} becomes
\beq
\lim_{\epsilon\rightarrow0}\EXP^{-\pi^2\alpha/(6r\omega_1\omega_2\epsilon)}\kappa^e(\alpha\epsilon)=\kappa^h(\alpha)\,,
\eeq
where for $|\re(\alpha)|<\re(\eta)$
\beq
\label{hkappar}
\kappa^h(\alpha)=\exp\left\{\int_0^\infty dx\left(-\frac{\alpha}{r\omega_1\omega_2 x^2}+\frac{\sinh(4\alpha x)\sinh(2r\eta x)}{2x\sinh(\omega_1rx)\sinh(\omega_2rx)\sinh(4\eta x)}\right)\right\}.
\eeq
This function satisfies the required functional relations
\beq
\kappa^h(\alpha)\kappa^h(-\alpha)=1\,,\quad\frac{\kappa^h(\eta-\alpha)}{\kappa^h(\alpha)}=\varphi_{r,0}(\ii(\eta-2\alpha))\,,
\eeq
for the result for the free energy \eqref{fzero} to hold, and the normalisation function $\kappa^h(\alpha)$ may be interpreted as the partition function per edge, for the lattice model with Boltzmann weights defined in the next subsection.  For $r=1$ the function represents the partition function per edge of both the Faddeev-Volkov model \cite{Bazhanov:2007mh,Bazhanov:2007vg}, and the generalisation of the latter based on the hyperbolic beta integral \cite{Spiridonov:2010em}. Further properties of $\kappa^h(\alpha)$ are summarised in Appendix \ref{app:functions}.

The following limit is also required
\beq
\lim_{\epsilon\rightarrow0}\EXP^{\pi^2(\omega_1+\omega_2)/(12r\omega_1\omega_2\epsilon)}(\p^{2r};\p^{2r})_\infty(\q^{2r};\q^{2r})_\infty=\frac{\pi}{r\epsilon\sqrt{\omega_1\omega_2}}\,,
\eeq
for the factors appearing in the one-spin Boltzmann weight $\s(\sigma_j)$ \eqref{ebws}.

\subsubsection{Star-triangle relation}

In this section the spins are now defined as
\beq
\label{hspindef}
\sigma_j=(x_j,m_j)\,,\quad 0\leq x_j<\infty\,,\quad m_j=0,1,\ldots,\floor{r/2}\,.
\eeq
The spectral parameters are restricted to the region $0<\alpha_i<\eta$, where the crossing parameter
\beq
\eta=(\omega_1+\omega_2)/2\,,
\eeq
is required to be real and positive valued.

The two-spin Boltzmann weights, are defined in terms of $\varphi_{r,m}(z)$, and $\kappa^h(\alpha)$, as
\beq
\label{hbw}
\w_\alpha(\sigma_i,\sigma_j)=\frac{1}{\kappa^h(\alpha)}\frac{\varphi_{r,m_i+m_j}(x_i+x_j+\ii\alpha)\,\varphi_{r,m_i-m_j}(x_i-x_j+\ii\alpha)}{\varphi_{r,m_i+m_j}(x_i+x_j-\ii\alpha)\,\varphi_{r,m_i-m_j}(x_i-x_j-\ii\alpha)}\,.
\eeq
The model is in a physical regime when $\omega_1=\omega^*_2$, where this Boltzmann weight is positive and real valued.  

The one-spin Boltzmann weight, is given by
\beq
\label{hbws}
\begin{array}{rcl}
\ds\s(\sigma_j)&=&\ds\frac{\varepsilon_j}{r\sqrt{\omega_1\omega_2}}\,\varphi_{r,-2m_j}(-2x_j-\ii\eta)\,\varphi_{r,2m_j}(2x_j-\ii\eta)\\[0.6cm]
&=&\ds\frac{4\varepsilon_j}{r\sqrt{\omega_1\omega_2}}\,\sinh\left(\frac{2\pi}{\omega_1r}(x_j-\ii\omega_1m_j)\right)\sinh\left(\frac{2\pi}{\omega_2r}(x_j+\ii\omega_2m_j)\right).
\end{array}
\eeq

The Boltzmann weight \eqref{hbw}, satisfies the spin reflection symmetry \eqref{spinrefl}, and  the following boundary conditions analogous to \eqref{msbc}
\beq
\begin{array}{rcl}
\ds\left.\w_\alpha(\sigma_i,\sigma_j)\right|_{\alpha=0}&=&\ds1,\\[0.3cm]
\ds\left.\w_{\eta-\alpha}(\sigma_i,\sigma_j)\right|_{\alpha\rightarrow0}&=&\ds\frac{\varepsilon_j}{\s(\sigma_i)}\left(\delta(x_i\!+\!x_j)\,\delta_{\llbracket m_i+m_j\rrbracket_r,0}+\delta(x_i\!-\!x_j)\,\delta_{\llbracket m_i-m_j\rrbracket_r,0}\right),
\end{array}
\eeq
where $\s(\sigma_i)\neq0$.

Formally the limit \eqref{hyplim1} of the elliptic star-triangle relation \eqref{str} gives the following new star-triangle relation
\beq
\label{hstr}
\begin{array}{r}
\ds\sum_{m_0=0}^{\floor{r/2}}\,\int^\infty_0\! dx_0\,\iS(\sigma_0)\,\iW_{\eta-\alpha_i}(\sigma_i,\sigma_0)\,\iW_{\eta-\alpha_j}(\sigma_j,\sigma_0)\,\iW_{\eta-\alpha_k}(\sigma_k,\sigma_0)\phantom{\,,}\\[0.4cm]
\ds=\iW_{\alpha_i}(\sigma_j,\sigma_k)\,\iW_{\alpha_j}(\sigma_i,\sigma_k)\,\iW_{\alpha_k}(\sigma_j,\sigma_i)\,,
\end{array}
\eeq
with Boltzmann weights defined in \eqref{hbw}, and \eqref{hbws}, and the spectral parameters satisfying $\eta=\alpha_i+\alpha_j+\alpha_k$.  The star-triangle relation \eqref{hstr} arises as a particular case of a hyperbolic hypergeometric sum/integral identity \eqref{hident}, derived in Appendix \ref{app:hyper}.  The interpretation of the star-triangle relation \eqref{hstr} in terms of dual $N=2$ $S_b^3/\mathbb{Z}_r$ partition functions, is described in the next section.

The partition function of the lattice model is given by \eqref{z-main}, with the Boltzmann weights defined in \eqref{hbw}, and \eqref{hbws}, and the integral and sum taken over all values of spins defined in \eqref{hspindef}.  As is expected for $r=1$, the star-triangle relation \eqref{hstr} reduces to a generalisation  \cite{Spiridonov:2010em} of the star-triangle relation for the Faddeev-Volkov model \cite{Bazhanov:2007mh,Bazhanov:2007vg}.  The limit $r\rightarrow\infty$ results in the ``rational'' solution of the star-triangle relation given in terms of the Euler gamma function \cite{Kels:2013ola,Kels:2015bda,Bazhanov:2007vg}, corresponding to the matching of dual two-dimensional $\mathcal{N}=(2,2)$ partition functions \cite{Benini:2011nc,Yamazaki:2013fva,Spiridonov2014}.

A new solution of the star-triangle relation is also expected to be obtained as a particular limit of \eqref{hstr}, when the real component of each spin in \eqref{hstr} is taken to infinity \cite{Spiridonov:2010em}, such that only a dependence on spin differences $x_i-x_j$, $m_i-m_j$ will remain.  For supersymmetric gauge theory, such a reduction corresponds to the breaking of the $SU(2)$ gauge group to $U(1)$.  The resulting identity is expected to be directly related to a particular self-dual solution of the star-triangle relation given in terms of a quantum dilogarithm with real and integer variables, which was recently obtained by Kashaev \cite{Kashaev:2015nya} from considerations of gauge invariance in quantum field theory.  The details of this calculation however are beyond the scope of this paper.

\section{The hyperbolic case from supersymmetric gauge theory}\label{sec:gaugetheory}


In this section we consider supersymmetric duality for three-dimensional $\mathcal N=2$ theories on the squashed lens spaces, which are free quotients of squashed three-sphere $S_b^3$ by $\mathbb{Z}_r$. Using the gauge/YBE correspondence we show that the star-triangle relation (\ref{hstr}) results from the invariance of the three-dimensional squashed lens partition functions under supersymmetric duality.

\subsection{Supersymmetric partition function on the squashed lens space} 

We start by defining the objects of interest, namely the general form of the three-dimensional $\mathcal N=2$ lens partition function, supersymmetric duality and then discuss the gauge/YBE correspondence. 

Besides the ordinary generators of the Poincare algebra the three-dimensional $\mathcal N=2$ supersymmetric algebra has four real supercharges. These theories have a gauge group $G$ and a global symmetry group $F$. The gauge group multiplets belong to the adjoint representation of $G$ whereas chiral multiplets belong to a suitable representation of $G$ and $F$ (in our case the fundamental). The supersymmetry algebra contains the $SO(2)$ R-symmetry which rotates supercharges. 

In recent years, there have been extensive studies on exactly calculable quantities of supersymmetric gauge theories in diverse dimensions due to the use of the supersymmetric localization technique  \cite{Nekrasov:2002qd,Pestun:2007rz}. This powerful analytical tool enables us to compute exact quantities\footnote{We will not discuss here the supersymmetric localization technique, since we will not use it except the fact that it provides exact results, the short review of the subject can be found in \cite{Hosomichi:2015jta}. The idea of localization was applied to three-dimensional $\mathcal N=2$ supersymmetric theories in \cite{Kapustin:2009kz}.} such as superconformal indices, partition functions on compact manifolds, Wilson loops, ’t Hooft loops, surface operators and so on. For our purposes such an exactly calculable quantity is the partition function on the squashed lens space $S_b^3/{\mathbb Z}_r$. This partition function has been first obtained in \cite{Benini:2011nc} and studied  in  \cite{Imamura:2012rq, Imamura:2013qxa,Nieri:2015yia,Nedelin:2016gwu}.

The squashed lens space $S_b^3/\mathbb{Z}$ is defined as the squashed three-sphere\footnote{The particular reason to consider supersymmetric theories on the squashed sphere is reproducing the quantities in Liouville theory which has the parameter $b$.}  
\begin{equation} 
S_b^3=\{(x,y)\in \mathbb{C}^2| \;\; b^2|x|^2+b^{-2}|y|^2=1\} \;,
\end{equation}
with the identification
\begin{equation} \label{squashed}
(x,y) \sim (e^{\frac{2\pi \ii}{r}} x,  e^{-\frac{2\pi \ii}{r}} y)~.
\end{equation}

As we mentioned, the localization technique enables us to calculate the partition function of $\mathcal{N}=2$ theories on the squashed lens space exactly\footnote{One can find the details of the computations, for instance in \cite{Alday:2012au} where authors consider a three-dimensional $\mathcal N=2$ Chern-Simons-matter theory on lens space $S_b^3/\mathbb{Z}_r$. Note that the localization technique in this case is quite similar to the partition function computations on $S_b^3$ \cite{Hama:2011ea}. It is also possible to derive the lens partition function from the four-dimensional index on $S^1 \times S^3/\mathbb{Z}_r$ via dimensional reduction \cite{Benini:2011nc,Yamazaki:2013fva} (see the next section).}, as a result the partition function is decomposed in the following form
\begin{align} \label{generalZ}
Z & =  \sum_{m}\int\frac{ \prod^{\text{rank G}} dz_j}{2\pi \ii \prod_k |\mathcal{W}_k|} \; Z_{\rm cl}[z, m] \; Z_{\rm vector}[z, m] \; Z_{\rm matter}[z, m] \;.
\end{align}
Here the sum is over the holonomies
\begin{equation}
m \ = \ \frac{r}{2 \pi} \int_C A_\mu dx^\mu \;,
\end{equation}
where the integration over a non-trivial cycle $\mathbb{C}$ on $S_b^3/{\mathbb Z}_r$ and $A_\mu$ is the gauge field.

The $z_j$ variables are associated with the Weyl weights for the Cartan subalgebra of the gauge group $G$. The $k$ is the rank of gauge group G and the prefactor $|W_k|$ is the order of the Weyl group of gauge group which is broken by holonomy into a product of $r$ subgroups
\begin{equation}
G\to \prod_{k=0}^{r-1} G_k \;.
\end{equation}
The classical term $Z_{\rm cl}$ is given by non-zero contributions from classical action of the Chern-Simons term and Fayet-Iliopoulos term. In our examples in the next sections we only discuss theories without the Chern-Simons terms, therefore $Z_{\rm cl}$ will be absent in our expressions.

There are two other contributions to the partition function, $Z_{\text{vector}}$ coming from vector multiplets and $Z_{\text{matter}}$ coming from the chiral multiplets. These contributions are as follows: the one-loop contribution of chiral multiplets\footnote{The chiral multiplet consists of a complex scalar field, a complex Dirac fermion and a complex auxiliary scalar field.} is given by
\begin{equation}
Z_{\rm matter}=
\prod_{i}\prod_{\rho_i}\prod_{\phi_i}
\hat s_{b,-\rho_i( m)-\phi_i( n)} 
\left(\ii\frac{Q}{2} (1-\Delta_i)-\rho_i(z)-\phi_i( \Phi)\right)~,
\end{equation}
where $i$ labels chiral multiplets,  $\rho_i,\phi_i$, are respectively the weights of the representation of the   gauge  and  flavor groups\footnote{Note that $\phi_i$ correspond to the real masses associated to flavor group.} and $\Delta_i$ the Weyl weight of $i$'th chiral multiplet. Here $Q=b+\frac{1}{b}$ with the squashing parameter\footnote{The squashing parameter can be real or a phase \cite{Imamura:2011wg, Closset:2012ru}.} $b^2=\omega_2/\omega_1$. The function  $\hat s_{b,-m}$ is the improved double sine function \cite{Nieri:2015yia}, defined here in terms of \eqref{hgammar} as
\begin{equation}
\label{sbdef}
\hat s_{b,-m}(x)=\sigma(m)\,\varphi_{r,m}(x)\,,
\end{equation}
where $\sigma$ is the following sign factor 
\begin{equation}
\label{signfactor}
\sigma(m)=\EXP^{\frac{\ii \pi}{2r}(\llbracket m\rrbracket(r-\llbracket m\rrbracket)-(r-1)m^2)}\,. 
\end{equation}
The double sine function (related to the $r=1$ case of \eqref{sbdef}) is a variant of Faddeev's non-compact quantum dilogarithm, which appears in various branches of mathematical physics.  Many properties of this function can be found in \cite{BultThesis}.

Note that it is actually a product representation of \eqref{sbdef}
\beq
\label{quantdilog}
\hat s_{b,-m}(x)=\sigma(m)\,\EXP^{\frac{\pi\ii}{2}B_\varphi(x,m,\omega_1,\omega_2)}\prod_{j=0}^{r-1}\frac{(\EXP^{2\pi(x+\ii\omega_2\llbracket m\rrbracket)/(\omega_2r)}\,(\EXP^{\pi\ii(\omega_1+\omega_2)/(\omega_2r)})^{2j+1};\EXP^{2\pi\ii\omega_1/(\omega_2r)})_\infty}{(\EXP^{2\pi(x-\ii\omega_1\llbracket m\rrbracket)/(\omega_1r)}\,(\EXP^{-\pi\ii(\omega_1+\omega_2)/(\omega_1r)})^{2j+1};\EXP^{-2\pi\ii\omega_2/(\omega_1r)})_\infty}\,,
\eeq
that naturally arises from gauge theory calculations, rather than the integral representation obtained from \eqref{hgammar}.  Here $B_\varphi$ is a particular combination of multiple Bernoulli polynomials defined in \eqref{bphidef}.  However when the squasing parameter is real, the infinite product representation (\ref{quantdilog}) is not valid, and one needs to use the integral representation obtained from (\ref{hgammar}).  An example of this is in the case of the usual round sphere, where one has to set $b = 1$.

The one-loop contribution of the vector multiplet\footnote{The vector multiplet consists of a gauge field, a complex Dirac fermion, a real scalar field and auxiliary scalar field.} for theory with non-abelian gauge group\footnote{For abelian vector fields, the one-loop determinant is trivial.}, combined with the Vandermonde determinant, is given by
\begin{align} \nonumber
Z_{\rm vector} & = \prod_{\alpha}\frac{1}{\hat s_{b, \alpha(m)} 
\left( \ii\frac{Q}{2}+ \alpha(z)\right)}\\ \label{3dvector}
& = \prod_{\alpha>0}4\sinh\frac{\pi}{r}\left(\frac{ \alpha(z)}{\omega_1} + \ii \alpha(m) \right)
\sinh\frac{\pi}{r}\left(\frac{ \alpha(z)}{\omega_2} - \ii \alpha(m) \right)~,
\end{align}
where the product is over the positive roots $\alpha$ of the gauge group $G$.

\subsection{\texorpdfstring{$3d$ ${\mathcal N}=2$} ~~supersymmetric duality}
\label{3d}

About two decades ago, Seiberg argued \cite{Seiberg:1994pq} a highly non-trivial statement about four-dimensional $\mathcal N=1$ supersymmetric gauge theories, that such theories with different ultraviolet behavior may flow to the same infrared fixed point, where these theories describe the same physics, {\it i.e.} an observer testing the low energy physics cannot distinguish the dual theories. The duality statement extends to other dimensions, particularly to three-dimensional $\mathcal N=2$ theories\footnote{Three-dimensional supersymmetric duality is first studied in \cite{Karch:1997ux,Aharony:1997gp}. Note that often four-dimensional dualities are called Seiberg duality, whereas three-dimensional $\mathcal N=2$ SQCD with $U(N)$ gauge group discussed in \cite{Aharony:1997gp} is called Aharony duality. Recently, three-dimensional Seiberg duality for $SU(N)$ gauge group is found in \cite{Aharony:2013dha} and special cases discussed in \cite{Gahramanov:2013rda,Gahramanov:2016wxi}.}.

Supersymmetric dualities have passed a number of consistency checks. An important test for the duality is an equivalence of partition functions of dual theories in the infrared fixed point\footnote{As usual it is not proof of the duality, but very solid evidence for it.}
\begin{equation}
    Z_{\text{theory A}} = Z_{\text{theory B}} \;.
\end{equation}
For several three-dimensional $\mathcal N=2$ dualities (mirror symmetry, Seiberg-like dualities {\it etc.}) such checks have been verified at the level of sphere partition functions (e.g. \cite{Kapustin:2010xq}), squashed sphere partition functions (e.g. \cite{Dolan:2011rp,Gahramanov:gka,Amariti:2015vwa}), superconformal indices (e.g. \cite{Krattenthaler:2011da,Kapustin:2011jm,Gahramanov:2013rda,Gahramanov:2016wxi}), lens partition functions (e.g. \cite{Imamura:2012rq,Imamura:2013qxa}) and so on.

Now let us consider the following three-dimensional $\mathcal N=2$ supersymmetric duality\footnote{The duality discussed here is a special case of $SP(2N)$ duality considered in \cite{Aharony:2013dha}.}: the \textbf{theory A} and its low-energy description \textbf{theory B}, which can be described purely in terms of composite gauge singlets\footnote{These theories are confining theories, i.e. all of the massless degrees of freedom are gauge (color) singlet particles}. 

\medskip

\begin{itemize} 

\item \underline{Theory A:} This theory has gauge group $SU(2)$ and the flavor group $SU(6)$. The field content is: chiral multiplets that transform under the fundamental representation of the gauge group and the flavor group, the vector multiplet that transform as the adjoint representation of the gauge group. The lens partition function reads
\begin{align} \nonumber
Z_{\rm SQCD} & = \sum_{m_0=0}^{r-1}\int_{\mathbb{R}}\frac{ dx_0}{r\sqrt{\omega_1\omega_2}}\;2\sinh\frac{2\pi}{r\omega_1}(x_0- \ii\omega_1 m_0)\sinh\frac{2\pi}{r\omega_2}(x_0+\ii\omega_2 m_0) \\ 
& \qquad  \qquad \qquad \times \prod_{k=1}^{6}\frac{\hat s_{b,-m_0-m_{k}}(x_0+x_{k}+\ii Q/2)}{\hat s_{b,-m_0+\bar m_{k}}(x_0-x_k-\ii Q/2)}~,
\end{align}
with the balancing condition $\ii\sum_{i=1}^6 x_i=Q$, $\sum_{i=1}^6 m_i=0$.
The first line of the partition function includes the contributions of a vector multiplet, while the second line contains the contribution of chiral multiplets.

\item \underline{Theory B:} In the dual description of the theory there is no gauge symmetry\footnote{One can see that dual theories have different gauge group, but same global symmetries. Indeed, gauge symmetry is redundancy of the theory, whereas the global symmetries are observables, hence cannot be different in two descriptions of the same theory.}, there are fifteen chiral multiplets in the totally antisymmetric tensor representation of the flavor group. The lens partition function of the theory is given by the simpler expression
\begin{equation}
Z \ = \ \prod_{1\leq j<k\leq 6} \hat s_{b,-m_{j}-m_{k}}(x_j+x_k+\ii Q/2) \;.
\end{equation}
Since all physical degrees of freedom of the theory B are gauge invariant (since it has no gauge group symmetry) there is no any summation (no holonomy) and integration in the expression of the partition function. 

\end{itemize}

Due to the supersymmetric duality one finds the equality of lens partition functions
\begin{align} \nonumber
& \sum_{m_0=0}^{r-1}\int_{\mathbb{R}}\frac{ dx_0}{r\sqrt{\omega_1\omega_2}}\;2\sinh\frac{2\pi}{r\omega_1}(x_0- \ii\omega_1 m_0)\sinh\frac{2\pi}{r\omega_2}(x_0+\ii\omega_2 m_0) \\ \label{pfidentity}
& \qquad  \qquad \qquad \times \prod_{k=1}^{6}\frac{\hat s_{b,-m_0-m_{k}}(x_0+x_{k}+\ii Q/2)}{\hat s_{b,-m_0+\bar m_{k}}(x_0-x_k-\ii Q/2)} \ \ =  \prod_{1\leq j<k\leq 6} \hat s_{b,-m_{j}-m_{k}}(x_j+x_k+\ii Q/2) \;,
\end{align}

with the balancing condition $\ii\sum_{i=1}^6 x_i=Q$, $\sum_{i=1}^6 m_i=0$.  In Appendix \ref{app:hyper}, details are given of the derivation of \eqref{pfidentity} as the limit of the elliptic hypergeometric sum/integral \cite{Kels:2015bda} corresponding to duality of lens indices of four-dimensional $\mathcal{N}=1$ supersymmetric gauge theory.

In \cite{Dolan:2011rp} it was noticed that the three-dimensional $\mathcal N=2$ squashed sphere partition functions have similar structure to four-dimensional $\mathcal N=1$ superconformal indices and one can obtain the latter via dimensional reduction.  Using the results of \cite{Dolan:2011rp}, Benini {\it et. al.} \cite{Benini:2011nc} described the procedure which reduces\footnote{Of course, the three-dimensional duality obtained from the four-dimensional duality differs from the naive dimensional reduction, see \cite{Aharony:2013dha} for details.} four-dimensional $\mathcal N = 1$ lens index to three-dimensional $\mathcal N = 2$ lens partition function. Geometrically, one needs to consider the four-dimensional $\mathcal N=1$ supersymmetric dual theories on a $S_b^3/\mathbb{Z}_r \times S^1$. Then shrinking the circle $S^1$ to zero gives rise to a three–dimensional supersymmetric theory with the same amount of supercharges on $S_b^3/\mathbb{Z}_r$. From the perspective of special functions this reduction is just the limit (\ref{hyplim2}) discussed in the previous section.

Note that the balancing conditions are imposed by the effective superpotential and the theories described above are dual only in the presence of certain superpotentials (for detail see e.g. \cite{Khmelnitsky:2009vc}). In order to obtain the correct three-dimensional duality via dimensional reduction one needs to add \cite{Aharony:2013dha} to theory A  the superpotential $W=\eta Y$ and to theory B the superpotential $W=\tilde{\eta}\tilde{Y}$, where $Y(\tilde{Y})$ is the low-energy limit of the monopole operator and $\eta(\tilde{\eta})$ is the dynamical scale of the corresponding four-dimensional theory (instanton factor).

The duality above is precisely the one considered in \cite{Teschner:2012em} where the authors presented the squashed sphere partition functions\footnote{In case when $r=1$, one obtains an identity for the squashed sphere partition functions \cite{Teschner:2012em}.} for dual theories. The proof of the integral identity for the squashed sphere partition functions can be found e.g., in \cite{BultThesis,Bult2007}. The matching of the superconformal indices for this duality was shown in \cite{Gahramanov:2016wxi} (see also \cite{Gahramanov:2015cva}).

As for the case $r=1$ \cite{Bult2007} we expect that the integral identity (\ref{pfidentity}) has the Weyl symmetry group of the exceptional root system $E_6$, for the case where each $m_i=0$.

\subsection{Integrability from duality}

In the context of gauge/YBE correspondence the spin lattice models of Section \ref{sec:newsolutions} can be identified with the quiver gauge theory with $SU(2)$ gauge groups. We associate a quiver diagram\footnote{Mathematically, a quiver is  a pair of $(V,E)$, where $V$ is a set of vertices (loops) and $E$ a set of arrows such that each arrow begins and ends on vertices, see, e.g. \cite{He:2004rn,Yamazaki:2008bt} for details.} to three-dimensional $\mathcal N=2$ supersymmetric gauge theory in the following way. On the vertices\footnote{In supersymmetry literature widely used loop instead of vertex.} we have gauge groups, in our case the corresponding gauge group of the theory is $\prod^{N} SU(2)$. The bifundamental matter content is represented as lines between gauge groups\footnote{In principle, one needs to specify fundamental and anti-fundamental representations by an arrow (outgoing and incoming arrows, respectively), but we do not need it here.}. The partition function of the corresponding integrable model is equivalent to the lens partition function of the corresponding supersymmetric quiver gauge theory with $SU(2)$ gauge groups. The contribution of chiral and vector multiplets to the lens partition function correspond to the Boltzmann weights for the nearest-neighbor and the self- interaction, respectively.  

The boundary conditions for the Boltzmann weights play a role of the ''chiral'' symmetry breaking phenomenon on the gauge theory side \cite{Spiridonov:2014cxa,Gahramanov:2015cva}. It is not well-studied subject on the level of supersymmetric partition functions and much work remains to be done in this direction.

\begin{figure}[tbh]
\centering
\begin{tikzpicture}[scale=2]

\begin{scope}[xshift=10]
\draw (-2,1)--(-2,0.18);
\draw (-2.87,-0.5)--(-2.16,0.01);
\draw (-1.13,-0.5)--(-1.82,0.01);
\draw (-2,0) circle [radius=0.17] node {$x_0$};
\draw (-1.84,1.35) rectangle ++(-0.35, -0.35) node[midway] {$x_i$};
\draw (-2.73,-0.5) rectangle ++(-0.35, -0.35) node[midway] {$x_j$};
\draw (-1.03,-0.5) rectangle ++(-0.35, -0.35) node[midway] {$x_k$};
\end{scope}


\fill[white!] (0.5,0.3) circle (0.01pt)
node[left=0.05pt] {\Large\color{black}$=$};

\begin{scope}[xshift=-10]
\draw[dashed,thick] (2.97,1.40) circle (0.26cm);
\draw[dashed,thick] (1.7,-0.85) circle (0.22cm);
\draw[dashed,thick] (4.2,-0.85) circle (0.22cm);

\node[anchor=base] at (3.04,1.06) (i) {};
\node[anchor=base] at (1.86,-0.55) (j) {};
\node[anchor=base] at (3.95,-0.55) (k) {};

\path[-] (i) edge (j);
\path[-] (i) edge (k);
\path[-] (k) edge (j);

\filldraw[fill=white,draw=black] (3.15,1.35) rectangle ++(-0.35, -0.35) node[midway] {$x_i$};
\filldraw[fill=white,draw=black] (2.03,-0.45) rectangle ++(-0.35, -0.35) node[midway] {$x_j$};
\filldraw[fill=white,draw=black] (4.2,-0.45) rectangle ++(-0.35, -0.35) node[midway] {$x_k$};
\end{scope}

\end{tikzpicture}
\caption{Seiberg duality: the boxes correspond to $SU(2)$ flavor subgroups
and the circle represents $SU(2)$ gauge group. The dash lines represent mesons giving contribution to the spin-independent $R$-factor in the star-triangle relation. In our case due to the renormalization of the Boltzmann weights the $R$-factor equals to one.}
\end{figure}
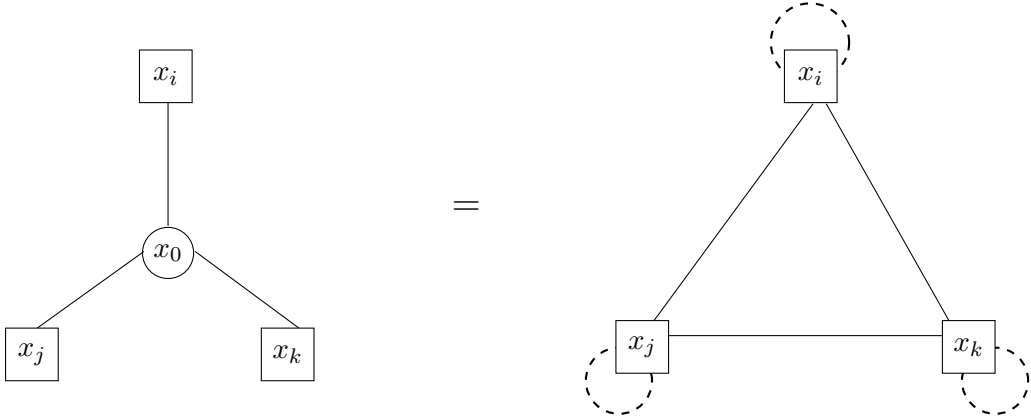

As already emphasized, the key observation is that the equality for the lens partition functions (\ref{pfidentity}) can be written as the star-triangle relation (\ref{hstr}). By adding a certain superpotential one may break flavor symmetry of both theories from $SU(6)$ group down to $SU(2) \times SU(2) \times SU(2)$. In fact, one can do it by introducing the following change of variables 
\beq 
\begin{array}{c}
\ds x_1=+x_i-\ii\alpha_i\,,\quad x_3=+x_j-\ii\alpha_j\,,\quad x_5=+x_k-\ii\alpha_k\,,\\[0.3cm]
\ds x_2=-x_i-\ii\alpha_i\,,\quad x_4=-x_j-\ii\alpha_j\,,\quad x_6=-x_k-\ii\alpha_k\,,
\end{array}
\eeq 
and
\beq 
m_1=m_i\,,\quad m_2=-m_i\,,\quad m_3=m_j\,,\quad m_4=-m_j\,,\quad m_5=m_k\,,\quad m_6=-m_k\,.
\eeq 
Under this change of variables the identity (\ref{pfidentity}) gets exactly the form of the star-triangle relation (\ref{hstr}).  


\section{The elliptic case from supersymmetric gauge theory}\label{sec:ellipticgauge}

\subsection{Lens index}

Here we briefly review the four-dimensional $\mathcal N=1$ lens supersymmetric index. 

The basic ingredients that are needed to know about four-dimensional $\mathcal N=1$ supersymmetric gauge theory are the following: it has a gauge group and a flavor symmetry group. The gauge group multiplets belong to the
adjoint representation of the gauge group whereas the chiral multiplets belong to a suitable representation of gauge and flavor group. The supersymmetry algebra contains the $U(1)$ $R$-symmetry which rotates supercharges.

The lens supersymmetric index was introduced in \cite{Benini:2011nc} and studied in \cite{Yamazaki:2013fva,Yamazaki:2013nra,Alday:2013rs, Nieri:2015yia}. The section will mainly follow the exposition in \cite{Yamazaki:2013nra}.

The four-dimensional $\mathcal N=1$ lens supersymmetric index is a generalization of the Witten index (partition function on $S^3/\mathbb{Z}_r \times S^1$) by including to the index symmetries of a theory commuting with a chosen supercharge. To construct the lens index let us consider, for example, the supercharges $Q$ which satisfy the following relation\footnote{for the full superconformal algebra, see e.g. \cite{Dolan:2008qi}.} 
\begin{equation} \label{comrel}
    \{Q,Q^{\dagger}\}=H-\frac32 R-2J_2 \;,
\end{equation}
where $H$, $R$ are the Hamiltonian in the radial quantization and the generator of the $R$-symmetry, respectively. $J_1$ and $J_2$ are the Cartan generators of the $SU(2) \times SU(2)$ isometry of $S^3$. Then one can define the lens supersymmetric index in the following way
\begin{equation}\label{Ind}
{\text{I} }(\{ t_i \},p,q)=\text{Tr}\left[ (-1)^F \EXP^{-\beta \{Q, Q^{\dagger}\}} p^{j_1+j_2+\frac{\mathfrak{r}}{2}} q^{-j_1+j_2+\frac{\mathfrak{r}}{2}} \prod_{i} t_i^{F_i} \right] \;,
\end{equation}
Here $(-1)^F$ is the fermion number operator, $F_i$ are generators of global symmetries commuting with $Q$ and $Q^{\dagger}$, and $t_i$ are the corresponding fugacities (additional regulators). The trace in the definition of the index is over the Hilbert space of the theory on a $S^{3}/\mathbb{Z}_r$. The states\footnote{Here we use eigenvalues of operators and therefore have different letters than in (\ref{comrel}).} with $E-\frac32 \mathfrak{r}-2j_2  \neq 0$ come in pairs and cancel out because of the factor $(-1)^F$, therefore the index is $\beta$--independent and counts states with $E-\frac32 \mathfrak{r}-2j_2 =0$. The index does not depend on coupling constants of the theory and it is invariant under marginal deformations of the theory. In the special case when $r=1$ one obtains the usual supersymmetric index \cite{Romelsberger:2005eg, Kinney:2005ej}. 

According to the Romelsberger prescription\footnote{Note that one can consider the lens index as a twisted partition function on $S_b^3/\mathbb{Z}_r \times S^1$. Then using localization technique one gets the same result for a twisted partition function.} \cite{Romelsberger:2005eg,Romelsberger:2007ec} for ${\mathcal N}=1$ theory with a weakly-coupled description one can write the lens index via the so-called ``plethystic'' exponential \cite{Feng:2007ur} of the single letter index. Then one can get the full index via the following integral over the gauge group\footnote{Since we are interested in gauge invariant physical observables.} (see the Appendix in \cite{Benini:2011nc} for details)
\begin{equation} \label{plethystic}
I(\{t_i\}, p, q; r)  = \sum_{m=0} \int d \mu_m(g)\, \exp \bigg ( \sum_{n=1}^{\infty}
\frac 1n \text{ind}\big(p^n ,q^n, \underline{z}^n , t_i^ n; m \big ) \bigg ),
\end{equation}
where $d \mu_m(g)$ is the gauge group--invariant Haar measure and $\text{ind}(p,q,z,t_i,m)$ stand for the index for single particle states. It is convenient to express the index as a product of contributions from chiral and vector multiplets
\begin{equation} 
I(\{t_i\}, p,q; r) = \sum_{m}
\int \frac{1}{|W|} Z_{gauge} (z_i,p,q; m)  \prod_{\Phi} Z_{\Phi}(z_i, t_a, p,q; m)  \prod_{i=1}^{\text{rank}G} \frac{dz_i}{2 \pi \ii z_i}.
\end{equation}
Here the sum is over holonomies $m$ on the $S^3$ and the prefactor $|W|= \prod_{m=0}^{r-1} ( \text{rank}\,G_m)!$ is the order of the Weyl group of $G$ which is ``broken'' by the holonomies into the product $G_0 \times G_1 \times \dots\times G_{r-1}$. The one loop superdeterminants of the vector and matter multiplets are expressed in terms of lens elliptic gamma functions $\Gamma_e(z,m;\sigma,\tau)$ defined in (\ref{legf2}).

\subsection{Duality}

Now one can consider the four-dimensional version of the duality from Section \ref{3d}, namely, supersymmetric duality states \cite{Seiberg:1994pq,Intriligator:1995ne} that the superconformal infrared fixed point of the following four-dimensional $\mathcal N=1$ theories are equivalent:

\begin{itemize}

\item \underline{Theory A}: $SU(2)$ gauge theory with $SU(6)$ flavor group\footnote{It is interesting to note that the global symmetry $SU(6)$ is enhanced to the exceptional symmetry group $E_6$ in the presence of the five-dimensional hypermultiplets \cite{Gahramanov:2013xsa} (see Section \ref{app:w(e)}).} with chiral multiplets forming the six dimensional fundamental representation. The lens supersymmetric index of this theory reads (see Appendix \ref{app:hyper} for the notations)
\begin{equation}
\ds \frac{(\p^r;\p^r)_\infty(\q^r;\q^r)_\infty}{2} \sum_{y=0}^{r-1}\int_{0}^{1}dz\,\frac{\prod_{i=1}^6\Gamma_e(t_i\pm z,u_i\pm y;\sigma,\tau)}{\Gamma_e(\pm 2z,\pm 2y;\sigma,\tau)}
\end{equation}
with the balancing conditions
\begin{equation}
\sum_{i=1}^6t_i= \sigma+\tau,\qquad\sum_{i=1}^6u_i=0\,,
\end{equation}
where $t_i$ and $z$ stand for the flavor fugacities and gauge groups, and $u_i$ and $y$ are corresponding holonomies associated with these groups, respectively. 

\medskip

\item \underline{Theory B}: without gauge degrees of freedom with chiral multiplets forming the 15-dimensional antisymmetric tensor representation. The lens supersymmetric index of the theory reads
\begin{equation}
\prod_{1\leq i<j\leq6}\!\!\Gamma_e(t_i+t_j,u_i+u_j;\sigma,\tau)\,.
\end{equation}

\end{itemize}

The identity of the lens supersymmetric indices\footnote{Note that the matching of superconformal indices for this duality, i.e. the $r=1$ case was shown by Dolan and Osborn in \cite{Dolan:2008qi}.} for dual theories has a form of the sum/integral identity (\ref{eident}) which can be written as the star-triangle relation (\ref{str}) \cite{Kels:2015bda}. All arguments about the gauge/YBE correspondence in this case structurally identical to the formal arguments of the Section \ref{3d}. 

\subsection{\texorpdfstring{$W(E_7)$} ~~symmetry and its breaking}\label{app:w(e)}

Let us consider the following integral elliptic hypergeometric sum/integral 
\beq
\label{Vfunc}
\ds V(\underline{t};q)=\frac{(\p^r;\p^r)_\infty(\q^r;\q^r)_\infty}{2} \sum_{y=0}^{r-1}\int_{0}^{1}dz\,\frac{\prod_{i=1}^8\Gamma_e(t_i\pm z,u_i\pm y;\sigma,\tau)}{\Gamma_e(\pm 2z,\pm 2y;\sigma,\tau)}\,,\ds
\eeq
with the balancing conditions
\beq
\sum_{i=1}^8 t_i= \sigma+\tau,\qquad\sum_{i=1}^8 u_i=0 \;,
\eeq
where the variables are
\beq
\p,\q,t_i\in\mathbb{C},\quad u_i\in\mathbb{Z},\quad |\p|,|\q|<1,\quad\im(t_i)>0\,,\quad i=1,\ldots,8\,.
\eeq

For the case $u_i=0, i=1, \; \ldots, 8$ this sum/integral obeys the $W(E_7)$-group of symmetries transformation -- the Weyl group of the exceptional root system of $E_7$, with the following transformation law (this property also was considered in \cite{rarified}) 
\begin{eqnarray}
V(\underline{t};q)=V(\underline{\tilde t};q)
\prod_{1\leq j<k\leq 4} \Gamma_e(t_j+t_k, 0;\sigma,\tau) \prod_{5\leq j<k\leq 8}  \Gamma_e(t_j+t_k, 0;\sigma,\tau)\,,
\label{E7trafo}
\end{eqnarray}
where 
\begin{equation}
\left\{
\begin{array}{cl}
\tilde t_j =t_j+ \epsilon ,&   j=1,2,3,4  \\
\tilde t_j = t_j-\epsilon , &    j=5,6,7,8
\end{array}
\right.;
\quad \epsilon=\frac{\sigma+\tau-t_1-t_2-t_3-t_4}{2}=\frac{t_5+t_6+t_7+t_8-\sigma-\tau}{2}.
\label{partrafo}\end{equation}
The proof is analogous to the proof in the usual elliptic case \cite{SpiridonovW(E7)} (see also \cite{Bult2007}). Note that the sum/integral (\ref{Vfunc}) can be put in the form of a star-star relation, namely it is a special case of the $R$-matrix found in \cite{Yamazaki:2013nra}. It also has a form of the the lens index of a four-dimensional ${\mathcal N}=1$ supersymmetric gauge theory with $SU(2)$ gauge group and $SU(8)$ flavor group; all matter multiplets in the fundamental representation of the gauge and flavor groups, vector multiplet is in the adjoint of the gauge group.

The lens sum/integral (\ref{Vfunc}) reduces to the lens elliptic beta sum/integral (\ref{eident}) when e.g. $t_7+t_8=\sigma+\tau$ and $u_7+u_8=0$. The latter has the $W(E_6)$ group of symmetry for the case when all holonomies associated with the flavor symmetries are absent (this property also appears to apply for a more general situation \cite{rarified}).  Since the lens elliptic gamma function \eqref{legf2} is $r$-periodic in the integer variable, the form of the hypergeometric equations satisfied by \eqref{Vfunc} should take a simpler form than in \cite{rarified}.  This is because of the different normalisation of \eqref{legf2} chosen in \cite{rarified}, details of this difference are given in Appendix \ref{app:modr}.

\section{Conclusion}

In this paper a new solution of the star-triangle relation \eqref{hstr} was given, that provides a generalisation of the Faddeev-Volkov model \cite{Bazhanov:2007vg,Bazhanov:2007mh,Spiridonov:2010em} to the case of both continuous and real valued spin components.  This new lattice model arises in the hyperbolic limit of the elliptic model previously obtained by the second author \cite{Kels:2015bda}.  The Boltzmann weights of the model \eqref{hbw},\eqref{hbws}, are given in terms of a generalisation of the hyperbolic gamma function \eqref{hgammar} (or non-compact quantum dilogarithm), that takes both integer and complex arguments.  The exact solution of the model is contained in the normalisation \eqref{hkappar} of Boltzmann weights, through an extension of the inversion relation method \cite{Stroganov:1979et,Zamolodchikov:1979ba,Baxter:1982xp,Bazhanov:2016ajm}.  The star-triangle relation of the model \eqref{hstr} was shown to arise as the duality of $S_b^3/\mathbb{Z}_r$ partition functions for three-dimensional supersymmetric gauge theory.  While the latter supersymmetric dualities are known in the literature \cite{Seiberg:1994pq,Intriligator:1995ne}, the approach here of studying them through the lens partition function and lens supersymmetric index is new.

There are many possible directions for future research.  For example, the Faddeev-Volkov model ($r=1$ case of Section \ref{sec:hyperbolicmodel}) notably has connections to discrete conformal geometry \cite{Bazhanov:2007vg}, and to classical discrete integrable equations in the quasi-classical limit \cite{Bazhanov:2007vg,Bazhanov:2016ajm}.  The Faddeev-Volkov model is also connected with the modular double of the quantum group $U_{\q}(sl_2)$ \cite{Faddeev:1999fe}, and with the lattice Liouville and sinh-Gordon models \cite{Faddeev:1993pe,Faddeev:1999fe,Faddeev:2000if,Kharchev:2001rs,Bytsko:2006ut}.  It would be interesting to see how these connections might be extended to the generalisation of the Faddeev-Volkov model considered here (for $r>1$).  It is also of interest to determine the hypergeometric properties \cite{BultThesis} of the hyperbolic sum/integral \eqref{hident} corresponding to the star-triangle relation \eqref{hstr}, as was recently done for the elliptic case \cite{rarified}.

Another possible future direction is to consider different limits or generalisations of the star-triangle relations \eqref{str}, \eqref{hstr}, in order to obtain new integrable models.  For example, many generalisations of the hyperbolic sum/integral \eqref{hident} are known for $r=1$ \cite{Rains2009}, it would be interesting to extend these results to the case $r>1$, and give the interpretation in terms of the gauge/YBE correspondence.  There are also expected to be many interesting limits of \eqref{hstr} (resp. \eqref{hident}), at the hyperbolic and rational levels.  For example, the limit $r \rightarrow \infty$ \cite{Yamazaki:2013fva} of \eqref{hstr}, is a rational limit that results in a lattice model with discrete and continuous spin variables \cite{Kels:2013ola,Kels:2015bda,Bazhanov:2007mh}, with Boltzmann weights given in terms of the Euler gamma function.  On the supersymmetry side this reduction (geometrically, the  $S^1$ fiber of the lens space $S_b^3/\mathbb{Z}_r$ shrinks to zero size and gives $S^2$) \cite{Benini:2011nc} gives the equality of the sphere partition functions of dual two-dimensional $\mathcal N = (2,2)$ supersymmetric gauge theories.  The root of unity limit could also be explored, which in the elliptic $r=1$ case was shown to reproduce well-known discrete spin integrable lattice models \cite{Bazhanov:2010kz}.

In the context of the gauge/YBE correspondence, it has recently been shown \cite{Maruyoshi:2016caf} that a surface defect in four-dimensional $\mathcal N=1$ supersymmetric gauge theory, corresponds to a transfer matrix constructed from Sklyanin's $L$-operators \cite{Sklyanin:1983ig} (see also \cite{rosengren2004sklyanin,Derkachov:2012iv}) of the lattice spin model.  It would be interesting to study this relationship for the cases in this paper.  Also, while the present paper provides a new example of the gauge/YBE correspondence in 2 dimensions, the question remains of whether it is possible to extend the gauge/YBE correspondence to 3 dimensions, where the analogue of the Yang-Baxter equation is Zamolodchikov's tetrahedron equation \cite{zamolodchikov1980tetrahedra}.  In one case, Gadde {\it et.al.} \cite{Gadde:2013lxa} have speculated that the superconformal indices of a two-dimensional $\mathcal N=(0,2)$ triality could provide a solution to the tetrahedron equation.  However apart from such speculation, a genuine extension of the correspondence to 3 dimensions is yet to be established.


\textbf{Acknowledgments}

The authors would like to thank Vyacheslav Spiridonov, and Masahito Yamazaki for helpful discussions on the subject.  IG would also like to thank Guman Garayev, Shahriyar Jafarzade, Gonenc Mogol, and Hjalmar Rosengren for fruitful discussions on the subject.  IG would like to thank the Institut des Hautes Études Scientifiques, IHES (Bures-sur-Yvette, France), Department of Mathematical Sciences, Chalmers University of Technology (Gothenburg, Sweden) and Nesin Mathematics Village (Izmir, Turkey), where some parts of the work was done for the warm hospitality. 

\begin{appendices}

\section{Properties of \texorpdfstring{$\varphi_{r,m}(z)$} ~~and \texorpdfstring{$\kappa^h(\alpha)$}.}\label{app:functions}

The following multiple Bernoulli polynomials \cite{Narukawa2004247,Barnes1904} are used throughout the appendices:
\beq
\label{bernoulli}
\begin{array}{rcl}
\ds B_{1,1}(z,\omega_1)&=&\ds\frac{z}{\omega_1}-\frac{1}{2}\,,\\[0.5cm]
\ds B_{2,2}(z,\omega_1,\omega_2)&=&\ds\frac{z^2}{\omega_1\omega_2}-\frac{(\omega_1+\omega_2)z}{\omega_1\omega_2}+\frac{\omega_1^2+\omega_2^2+3\omega_1\omega_2}{6\omega_1\omega_2}\,,\\[0.5cm]
\ds B_{3,3}(z,\omega_1,\omega_2,\omega_3)&=&\ds\frac{z^3}{\omega_1\omega_2\omega_3}-\frac{3(\omega_1+\omega_2+\omega_3)z^2}{2\omega_1\omega_2\omega_3}+\frac{(\omega_1^2+\omega_2^2+\omega_3^2+3(\omega_1\omega_2+\omega_2\omega_3+\omega_3\omega_1))z}{2\omega_1\omega_2\omega_3}\\[0.5cm]
&&\ds-\frac{(\omega_1+\omega_2+\omega_3)(\omega_1\omega_2+\omega_2\omega_3+\omega_3\omega_1)}{4\omega_1\omega_2\omega_3}\,,
\end{array}
\eeq
for complex variables $z\in\mathbb{C}$, and $\omega_1,\omega_2,\omega_3\in\mathbb{C}-\{0\}$.

In the following, $r\in\{1,2,\ldots\}$, the two complex parameters $\omega_1,\omega_2$, satisfy $\re(\omega_1),\re(\omega_2)>0$, and the following definitions are used:
\beq
\begin{array}{c}
\ds\omega=\EXP^{\pi\ii/r}\,,\;\q=\EXP^{\pi\ii\omega_1/(\omega_2r)}\,,\;\qt=\EXP^{-\pi\ii\omega_2 /(\omega_1r)}\,,\;\qb=\EXP^{-\pi\ii r\omega_2/(2\eta)}\,,\\[0.4cm]
\ds\eta=(\omega_1+\omega_2)/2\,,\quad\llbracket m\rrbracket=m\mbox{ mod }r\,,\quad(x;\p)_\infty=\prod_{j=0}^\infty(1-x\p^j)\,,
\end{array}
\eeq
where $\llbracket m\rrbracket\in\{0,1,\ldots,r-1\}$, and $|\p|<1$.
\\[0.2cm]
{\bf The function $\varphi_{r,m}(z)$.} The function $\varphi_{r,m}(z)$ depends on an integer $m\mbox{ (mod }r\mbox{)}$, a complex variable $z$, and implicitly on the two complex parameters $\omega_1,\omega_2$. The properties for the $r=1$ case have been studied in various different forms, such as the hyperbolic gamma function \cite{Ruijsenaars:1997:FOA,BultThesis}, the non-compact quantum dilogarithm \cite{Faddeev:1994fw,Faddeev:2000if,Bazhanov:2007mh}, and the double sine function \cite{Barnes:1901,shintani1976,kurokawa1991,Ponsot:2000mt,Kharchev:2001rs}.  The latter functions only depend on the complex variables $z,\omega_1,\omega_2$, and are related to each other by a simple change of variables \cite{Spiridonov:2010em}.  Below the relevant properties are given for the function $\varphi_{r,m}(z)$, for the case of general $r\geq1$, which may be derived through straightforward generalisations of the $r=1$ case.
\\[0.4cm]
{\it (i) Definition:} For
\beq
-\re(\eta)-\min\left(\re(\omega_1)(r-\llbracket m\rrbracket),\re(\omega_2)\llbracket m\rrbracket\right)<\im(z)<\re(\eta)+\min\left(\re(\omega_1)\llbracket m\rrbracket,\re(\omega_2)(r-\llbracket m\rrbracket)\right),
\eeq
\beq
\label{defphi}
\begin{array}{rcl}
\ds\varphi_{r,m}(z)&=&\ds\exp\left\{\int_0^\infty\frac{dx}{x}\left(\frac{\ii z}{\omega_1\omega_2 rx}-\frac{\sinh(2x(\ii z-(\frac{r}{2}-\llbracket m\rrbracket)\omega_1))}{2\sinh(\omega_1rx)\sinh(2\eta x)}-\frac{\sinh(2x(\ii z+(\frac{r}{2}-\llbracket m\rrbracket)\omega_2))}{2\sinh(\omega_2rx)\sinh(2\eta x)}\right)\!\right\}\\[0.6cm]
&=&\ds\varphi_{1,0}(z+\ii\omega_1(r-2\llbracket m\rrbracket)/2;\omega_1r,2\eta)\,\varphi_{1,0}(z-\ii\omega_2(r-2\llbracket m\rrbracket)/2;\omega_2r,2\eta)\,,
\end{array}
\eeq
where
\beq
\varphi_{1,0}(z;\omega_1,\omega_2)=\exp\left\{\int^\infty_0\frac{dx}{x}\left(\frac{\ii z}{\omega_1\omega_2x}-\frac{\sinh(2\ii zx)}{2\sinh(\omega_1x)\sinh(\omega_2x)}\right)\right\}.
\eeq
\\[0.4cm]
{\it (ii) Functional equations:}
\beq
\label{feq1}
\varphi_{r,m}(z)\varphi_{r,-m}(-z)=1\,,
\eeq
~
\beq
\label{feq2}
\frac{\varphi_{r,m+1}(z+\ii\omega_1)}{\varphi_{r,m}(z)}=\frac{\ii}{2\sinh\left(\frac{\pi}{\omega_2r}(z+\ii(\eta+\omega_2\llbracket m\rrbracket))\right)},
\eeq
and \phantom{\eqref{feq2}}
\beq
\label{feq3}
\frac{\varphi_{r,m+1}(z-\ii\omega_2)}{\varphi_{r,m}(z)}=2\ii\sinh\left(\frac{\pi}{\omega_1r}(z-\ii(\eta+\omega_1\llbracket m\rrbracket))\right).
\eeq
\\[0.0cm]
{\it (iii) Poles and zeros:}
\beq
\label{zeropole}
\begin{array}{l}
\ds Poles:\;\left\{+\ii\eta+\ii\omega_2j+\ii\omega_1\left(rk+\llbracket m+j\rrbracket\right)\right\},\\[0.3cm]
\ds Zeros:\;\left\{-\ii\eta-\ii\omega_1j-\ii\omega_2\left(rk+\llbracket m+j\rrbracket\right)\right\},
\end{array}
\eeq
\\[0.0cm]
where $j,k=0,1,\ldots$.
\\[0.3cm]
{\it (iv) Product representation:} For $\im(\omega_1/\omega_2)>0$,
\beq
\label{phiprod}
\ds\varphi_{r,m}(z)=\ds\EXP^{\frac{\pi\ii}{2}B_\varphi(z,m,\omega_1,\omega_2)}\prod_{j=0}^{r-1}\frac{(\EXP^{2\pi(z+\ii\omega_2\llbracket m\rrbracket)/(\omega_2r)}\,(\q\omega)^{2j+1};\q^{2r})_\infty}{(\EXP^{2\pi(z-\ii\omega_1\llbracket m\rrbracket)/(\omega_1r)}\,(\qt/\omega)^{2j+1};\qt^{2r})_\infty}\,,
\eeq
where $B_\varphi(z,m,\omega_1,\omega_2)$ is defined as
\beq
\label{bphidef}
B_\varphi(z,m,\omega_1,\omega_2):=B_{2,2}(\ii z+\omega_1\llbracket m \rrbracket+\eta,r\omega_1,2\eta)+B_{2,2}(\ii z+\omega_2(r-\llbracket m\rrbracket)+\eta,r\omega_2,2\eta)\,.
\eeq
For $r=1$ this simplifies to
\beq
B_\varphi(z,m,\omega_1,\omega_2)=B_{2,2}(\ii z+\eta,\omega_1,\omega_2)\,.
\eeq
\\[0.0cm]
The above functional equations \eqref{feq1}-\eqref{feq3} can be derived directly from the definition \eqref{defphi} by using only elementary trigonometric identities.  Through the functional identities, the function $\varphi_{r,m}(z)$ may be continued to a meromorphic function on $z\in\mathbb{C}$, with the poles and zeroes \eqref{zeropole} lying entirely in the upper and lower half planes respectively.

The product representation \eqref{phiprod} may be obtained by expanding the integral in \eqref{defphi} as a sum over residues in the upper half plane.  Through the product representation the above hyperbolic gamma function can be put in the form of a generalisation of the double sine functions
\beq
\label{dblsine}
\varphi_{r,m}(\ii(z-\eta))=\EXP^{\frac{\pi\ii}{2}B_\varphi(\ii(z-\eta),m,\omega_1,\omega_2)}\prod_{j=0}^{r-1}\frac{(\EXP^{2\pi\ii(z+\omega_2\llbracket m\rrbracket)/(\omega_2r)}\,(\q\omega)^{2j};\q^{2r})_\infty}{(\EXP^{2\pi\ii(z-\omega_1\llbracket m\rrbracket)/(\omega_1r)}\,(\qt/\omega)^{2j+2};\qt^{2r})_\infty}.
\eeq
The representation \eqref{dblsine} is equivalent to the function $S_{2,m}(z)$, previously studied in the context of three-dimensional $\mathcal{N}=2$ partition functions on $S_b^3/\mathbb{Z}^3$ \cite{Nieri:2015yia}.  The product representation however is not valid for $\im(\omega_1/\omega_2)=0$, which includes part of the physical regime of the lattice model in Section \ref{sec:hyperbolicmodel} (when $\im(\omega_1)=0$, and $\omega_2=\omega_1$), and also includes the case when the squashing parameter $b^2$ is on the real line, for the supersymmetric gauge theory described in Section \ref{3d}.
\\[0.3cm]
{\bf The function $\kappa^h(\alpha)$.}  The function $\kappa^h(\alpha)$ represents the partition function per edge of the lattice model defined in Section \ref{sec:hyperbolicmodel}.  For $r=1$, the function $\kappa^h(\alpha)$ appeared\footnote{Specifically, for $r=1$, $\omega_1=b$, $\omega_2=1/b$, the function $\Phi(2\ii\alpha)$ in \cite{Bazhanov:2007vg,Bazhanov:2007mh} , is equivalent to $\EXP^{-\frac{\pi\ii}{2}B_\kappa(\alpha,b,1/b)}\kappa^h(\alpha)$, ($B_\kappa$ is defined in \eqref{bkappadef}). For $r=1$, and $\omega_3=\omega_1+\omega_2$, the function $m(\alpha)$ in \cite{Spiridonov:2010em} is equivalent to $\kappa^h(\alpha)$.} as the partition function per edge of both the Faddeev-Volkov model \cite{Bazhanov:2007mh,Bazhanov:2007vg}, and its generalisation based on the hyperbolic beta integral \cite{Spiridonov:2010em}, and has also appeared with respect to the quantum sinh-Gordon model \cite{Lukyanov:1996jj}.  The properties of $\kappa^h(\alpha)$ for general $r\geq1$ are summarised below, and may be obtained analogously to the properties of $\varphi_{r,m}(z)$.
\\[0.5cm]
{\it (i) Definition:} For $|\re(\alpha)|<\re(\eta)$,
\beq
\kappa^h(\alpha)=\exp\left\{\int_0^\infty \frac{dx}{x}\left(-\frac{\alpha}{r\omega_1\omega_2 x}+\frac{\sinh(4\alpha x)\sinh(2r\eta x)}{2\sinh(\omega_1rx)\sinh(\omega_2rx)\sinh(4\eta x)}\right)\right\}.
\eeq
\\[0.0cm]
{\it (ii) Functional equations:}
\beq
\kappa^h(\alpha)\kappa^h(-\alpha)=1\,,\qquad\frac{\kappa^h(\eta-\alpha)}{\kappa^h(\alpha)}=\varphi_{r,0}(\ii(\eta-2\alpha))\,.
\eeq
\\[0.0cm]
{\it (iii) Poles and zeros:}
\beq
\begin{array}{l}
\ds Poles:\;\left\{\!\!
\begin{array}{ll}
\ds\left\{+\eta\,(2j_1+1)+\frac{k_1\omega_1r}{2}+\frac{k_2\omega_2r}{2}\,,\;k_1+k_2-\left| k_1-k_2\right|=0\mbox{ mod }4\right\},&\ds r\mbox{ odd}\,,\\[0.3cm]
\ds\left\{+\eta\,(2j_2+1)+\frac{k_1\omega_1r}{2}+\frac{k_2\omega_2r}{2}\right\},&\ds r\mbox{ even}\,,
\end{array}\right.
\end{array}
\eeq
\beq
\begin{array}{l}
\ds Zeros:\;\left\{\!\!
\begin{array}{ll}
\ds\left\{-\eta\,(2j_1+1)-\frac{k_1\omega_1r}{2}-\frac{k_2\omega_2r}{2}\,,\;k_1+k_2-\left| k_1-k_2\right|=0\mbox{ mod }4\right\},&\ds r\mbox{ odd}\,,\\[0.3cm]
\ds\left\{-\eta\,(2j_2+1)-\frac{k_1\omega_1r}{2}-\frac{k_2\omega_2r}{2}\right\},&\ds r\mbox{ even}\,,
\end{array}\right.
\end{array}
\eeq
where $j_1=0,1,\ldots,r-1$, $j_2=0,1,\ldots,r/2-1$, and $k_1,k_2=0,1,\ldots$.
\\[0.3cm]
{\it (iv) Product representation:} For $\im(\omega_1/\omega_2)>0$, and $r$ odd,
\beq
\kappa^h(\alpha)=\EXP^{\frac{\pi\ii}{2} B_{\kappa}(z,\omega_1,\omega_2)}\frac{(\EXP^{\ii\pi\alpha/\eta}\,\qb;\qb^2)_\infty}{(-\EXP^{\ii\pi\alpha/\eta}\,\qb;\qb^2)_\infty}\prod_{j=0}^{r-1}\frac{(\EXP^{4\ii\pi\alpha/(\omega_2r)}(\q\omega)^{4j+2};\q^{4r})_\infty}{(\EXP^{4\ii\pi\alpha/(\omega_1r)}(\qt/\omega)^{4j+2};\qt^{4r})_\infty}\,.
\eeq
For $\im(\omega_1/\omega_2)>0$, and $r$ even,
\beq
\kappa^h(\alpha)=\EXP^{\frac{\pi\ii}{2} B_{\kappa}(z,\omega_1,\omega_2)}\prod_{j=0}^{r/2-1}\frac{(\EXP^{4\ii\pi\alpha/(\omega_2r)}(\q\omega)^{4j+2};\q^{2r})_\infty}{(\EXP^{4\ii\pi\alpha/(\omega_1r)}(\qt/\omega)^{4j+2};\qt^{2r})_\infty}\,,
\eeq
where $B_{\kappa}(z,\omega_1,\omega_2)$ is defined as
\beq
\label{bkappadef}
B_\kappa(z,\omega_1,\omega_2):=B_{2,2}(2(\eta+z),r\omega_1,4\eta)+B_{2,2}(2(\eta-z),r\omega_2,4\eta)\,.
\eeq

\section{Hyperbolic limit}
\label{app:hyper}

This section expands on the results of Section \ref{sec:hyperbolicmodel}, by providing details of the hyperbolic limit of the elliptic hypergeometric integral/sum identity \cite{Kels:2015bda} that corresponds to the star-triangle relation \eqref{str}.

Define the following elliptic nomes
\beq
\label{nomedef}
\p=\EXP^{2\ii\pi\sigma}\,,\quad\q=\EXP^{2\ii\pi\tau}\,,\quad\im(\sigma),\;\im(\tau) >0\,,
\eeq
and define the following combinations of the multiple Bernoulli polynomial $B_{3,3}(z;\sigma,\tau)$ \eqref{bernoulli}
\beq
R(z;\sigma,\tau)=\frac{B_{3,3}(z;\sigma,\tau,-1)+B_{3,3}(z-1;\sigma,\tau,-1)}{12}\,,
\eeq
\beq
\label{r2def}
\begin{array}{rcl}
\ds R_2(z,m;\sigma,\tau)&=&\ds R(z+m\sigma;r\sigma,\sigma+\tau)+R(z+(r-m)\tau;r\tau,\sigma+\tau)\\[0.3cm]
&=&\ds\frac{(\sigma+\tau-2z)(2z^2-2z(\sigma+\tau)+\sigma\tau(r^2+6(m-r)m)+1)}{24r\sigma\tau}\\[0.5cm]
&&\ds-\frac{(\sigma-\tau)(2m-r)(m-r)m}{12r}\,,
\end{array}
\eeq
where $z\in\mathbb{C}$, $m\in\mathbb{Z}$, $r\in \{1,2,\ldots\}$.  The second equality explicitly shows that $R_2(z,m;\sigma,\tau)$ is defined for $\sigma=-\tau$, and this case will be utilised below.

The lens elliptic gamma function is defined here as\footnote{This definition is related to \eqref{legf} by $\Phi(\pi((\sigma+\tau)/2-z),m)=\EXP^{-\phi_e(z,\llbracket m\rrbracket;\sigma,\tau)}\Gamma_e(z,\llbracket m\rrbracket;\sigma,\tau)$.}
\beq
\label{legf2}
\Gamma_e(z,m;\sigma,\tau)=\EXP^{\phi_e(z,\llbracket m\rrbracket;\sigma,\tau)}\gamma_e(z,\llbracket m\rrbracket;\sigma,\tau)\,,
\eeq
where
\beq
\label{littlegamma}
\gamma_e(z,m;\sigma,\tau)=\prod_{j,k=0}^\infty\frac{1-\EXP^{-2\pi\ii z}\p^{-m}(\p\q)^{j+1}\p^{r(k+1)}}{1-\EXP^{2\pi\ii z}\p^{m}(\p\q)^j\p^{rk}}\frac{1-\EXP^{-2\pi\ii z}\q^{-r+m}(\p\q)^{j+1}\q^{r(k+1)}}{1-\EXP^{2\pi\ii z}\q^{r-m}(\p\q)^j\q^{rk}}\,,
\eeq
and
\beq
\label{ellnorm}
\begin{array}{rcl}
\ds\phi_e(z,m;\sigma,\tau)&=&\ds 2\pi\ii\left(R_2(z,0;\sigma-1/2,\tau+1/2)-R_2(z,m;\sigma-1/2,\tau+1/2)\right)\\[0.3cm]
&=&\ds 2\pi\ii\left(R_2(z,0;\sigma,\tau)+R_2(0,m,1/2,-1/2)-R_2(z,m;\sigma,\tau)\right).
\end{array}
\eeq
The lens elliptic gamma function \eqref{legf2}, may also be written as the following product of two regular elliptic gamma functions
\beq
\label{legfprod}
\Gamma_e(z,m;\sigma,\tau)=\EXP^{\phi_e(z,\llbracket m\rrbracket;\sigma,\tau)}\,\Gamma_{e,1}(z+\sigma\llbracket m \rrbracket;r\sigma,\sigma+\tau)\,\Gamma_{e,1}(z+\tau(r-\llbracket m \rrbracket);r\tau,\sigma+\tau)\,,
\eeq
where \cite{Ruijsenaars:1997:FOA,Felder}
\beq
\label{egf2}
\begin{array}{rcll}
\ds\Gamma_{e,1}(z;\sigma,\tau)&=&\ds\prod_{j,k=0}^\infty\frac{1-\EXP^{-2\pi\ii z}\p^{j+1}\q^{k+1}}{1-\EXP^{2\pi\ii z}\p^j,\q^k}\,,&\ds\quad z\in\mathbb{C} \\[0.6cm]
&=&\ds\exp\left\{G_{e,1}(z;\sigma,\tau)\right\},&\ds\quad 0<\im(z)<\im(\sigma+\tau)\,,
\end{array}
\eeq
and
\beq
G_{e,1}(z;\sigma,\tau)=-\frac{i}{2}\sum_{k=1}^\infty\frac{\sin(k\pi(2z-\sigma-\tau))}{k\sin(k\pi\sigma)\sin(k\pi\tau)}\,.
\eeq
Note that the following sum of two $R_2$ polynomials may be explicitly written as
\beq
\begin{array}{l}
\ds2\pi\ii\left(R_2(z,0;\sigma,\tau)-R_2(z,m;\sigma,\tau)\right)=\frac{\pi\ii\,m(m-r)}{2r}\left(2z-(\sigma+\tau)+(\sigma-\tau)(2m-r)/3\right),
\end{array}
\eeq
and thus the normalisation factor in \eqref{legf2} is
\beq \label{zpc}
\phi_e(z,\llbracket m\rrbracket;\sigma,\tau)=\frac{\pi\ii\,\llbracket m\rrbracket\llbracket -m\rrbracket}{2r}\left(\sigma+\tau-2z+(1+\tau-\sigma)(\llbracket m\rrbracket -\llbracket-m\rrbracket)/3\right),
\eeq
which is the same normalisation that was previously used \cite{Kels:2015bda}, up to a simple rescaling of $z,\sigma,\tau$. Note that in the context of the lens supersymmetric index the factor (\ref{zpc}) is the contribution to the zero point energy of chiral and vector multiplets \cite{Benini:2011nc}.

The following useful compact notation for products of gamma functions will be used
\beq
\label{comnot}
\Gamma(z_1\pm z_2,u_1\pm u_2;\sigma,\tau)=\Gamma(z_1+z_2,u_1+u_2;\sigma,\tau)\,\Gamma(z_1-z_2,u_1-u_2;\sigma,\tau)\,,
\eeq
for $z_1,z_2\in\mathbb{C}$, and $u_1,u_2\in\mathbb{Z}$.

The lens elliptic gamma function \eqref{legf2} satisfies the following elliptic hypergeometric sum/integral identity \cite{Kels:2015bda}
\beq
\label{eident}
\ds\lambda\sum_{y=0}^{\floor{r/2}}\varepsilon(y)\int_{-\frac{1}{2}}^{\frac{1}{2}}dz\,\frac{\prod_{i=1}^6\Gamma_e(t_i\pm z,u_i\pm y;\sigma,\tau)}{\Gamma_e(\pm 2z,\pm 2y;\sigma,\tau)}\ds=\!\!\!\!\prod_{1\leq i<j\leq6}\!\!\Gamma_e(t_i+t_j,u_i+u_j;\sigma,\tau)\,,
\eeq
where
\beq
\label{lambdadef}
\lambda=\frac{(\p^r;\p^r)_\infty(\q^r;\q^r)_\infty}{2}\,,\qquad\qquad
\varepsilon(y)=\left\{
\begin{array}{ll}
1& \quad y=0\mbox{ or }\frac{r}{2}\,,\\[0.3cm]
2& \quad\mbox{otherwise}\,,
\end{array}
\right.
\eeq
and the variables are
\beq
\p,\q,t_i\in\mathbb{C},\quad u_i\in\mathbb{Z},\quad |\p|,|\q|<1,\quad\im(t_i)>0\,,\quad i=1,\ldots,6\,,
\eeq
and satisfy
\beq
\sum_{i=1}^6t_i= \sigma+\tau,\qquad\sum_{i=1}^6u_i=0\,.
\eeq
The truncation of the sum from $0\leq y\leq r-1$, to $0\leq y\leq \floor{r/2}$, follows from the invariance of the integral under the change of variables $z\rightarrow-z$, $y\rightarrow r-y$, and the inclusion of the factor $\varepsilon(y)$ correctly counts the integrals after employing this symmetry.  The star-triangle relation \eqref{str} is obtained from \eqref{eident} for a particular choice of variables $u_i,t_i$ \cite{Kels:2015bda}, and \eqref{eident} is equivalent to Spiridonov's elliptic beta integral \cite{SpiridonovEBF} for $r=1$.

In the hyperbolic limit
\beq
\label{hyplim}
\p=\EXP^{2\pi\ii\omega_1\epsilon}\,,\quad\q=\EXP^{2\pi\ii\omega_2\epsilon}\,,\quad \epsilon\rightarrow0^+\,,
\eeq
where $\im(\omega_1),\im(\omega_2)>0$, it will be shown that \eqref{eident} reduces to the following identity
\beq
\label{hident}
\frac{1}{2r\sqrt{-\omega_1\omega_2}}\sum_{y=0}^{\floor{r/2}}\varepsilon(y)\int^{\infty}_{-\infty}\!dz\,\frac{\prod_{i=1}^6\Gamma_h(t_i\pm z,u_i\pm y;\omega_1,\omega_2)}{\Gamma_h(\pm 2z,\pm 2y;\omega_1,\omega_2)}\ds=\!\!\!\!\prod_{1\leq i<j\leq6}\!\!\Gamma_h(t_i+t_j,u_i+u_j;\omega_1,\omega_2)\,.
\eeq
Here the function $\Gamma_h(z,m;\omega_1,\omega_2)$, is defined in terms of $\varphi_{r,m}(z;\omega_1,\omega_2)$ \eqref{defphi} as
\beq
\label{lhgf}
\begin{array}{rcl}
\ds\Gamma_h(z,m;\omega_1,\omega_2)&=&\ds\EXP^{\phi_h(\llbracket m\rrbracket)}\,\varphi_{r,m}(-z+(\omega_1+\omega_2)/2;-\ii\omega_1,-\ii\omega_2)\\[0.3cm]
&=&\ds\EXP^{\phi_h(\llbracket m\rrbracket)}\gamma_h(z,m;\omega_1,\omega_2)\,,
\end{array}
\eeq
where
\beq
\label{hypnorm}
\phi_h(m)=2\pi\ii R_2(0,m,1/2,-1/2)\,,
\eeq
and
\beq
\label{littlegammah}
\gamma_h(z,m;\omega_1,\omega_2)=\Gamma_{h,1}(z+\omega_1\llbracket m\rrbracket;r\omega_1,\omega_1+\omega_2)\,\Gamma_{h,1}(z+\omega_2(r-\llbracket m\rrbracket);r\omega_2,\omega_1+\omega_2)\,.
\eeq
The $\Gamma_{h,1}(z;\omega_1,\omega_2)$ is the usual hyperbolic gamma function \cite{Ruijsenaars:1997:FOA}, defined for $0<\im(z)<\im(\omega_1+\omega_2)$ as
\beq
\label{hgf}
\Gamma_{h,1}(z;\omega_1,\omega_2)=\EXP^{G_{h,1}(z;\omega_1,\omega_2)}\,,
\eeq
where
\beq
G_{h,1}(z;\omega_1,\omega_2)=\frac{\ii}{2}\int_{0}^{\infty}\!\frac{dx}{x}\left(\frac{2z-\omega_1-\omega_2}{\pi\omega_1\omega_2\,x}-\frac{\sin(\pi(2z-\omega_1-\omega_2)x)}{\sin(\pi\omega_1x)\sin(\pi\omega_2x)}\right).
\eeq
The products of the $\Gamma_h$ in \eqref{hident} follow the convention given in Equation \eqref{comnot}, and the variables are
\beq
\omega_1,\omega_2,t_i\in\mathbb{C}\,,\quad u_i\in\mathbb{Z}\,,\quad\im(\omega_1),\im(\omega_2),\im(t_i)>0\,,\quad i=1,\ldots,6\,,
\eeq
and satisfy
\beq
\sum_{i=1}^6t_i=\omega_1+\omega_2,\quad\sum_{i=1}^6u_i=0\,.
\eeq
Equation \eqref{hident} is basically a generalisation of the univariate hyperbolic beta integral \cite{STOKMAN2005119}, with 6 integer variables in addition to the 6 complex variables, and is equivalent to the hyperbolic beta integral for $r=1$.  

Rather than directly substituting the hyperbolic limits of \eqref{legf2} into the integral \eqref{eident} as was done in Section \ref{sec:hyperbolicmodel}, asymptotic estimates that were previously obtained by Rains \cite{Rains2009} will be used in a relevant form for the gamma functions \eqref{legf2}, and \eqref{lhgf}, to show the convergence of the elliptic identity \eqref{eident} to the hyperbolic identity \eqref{hident}.   Obtaining relevant asymptotic estimates for the gamma functions \eqref{legf2}, and \eqref{lhgf} is rather straightforward, since both gamma functions may be written in terms of products of two standard elliptic, and hyperbolic gamma functions respectively, for which the previous asymptotic results \cite{Rains2009} apply.   Specifically, the results here provide an extension to values $r>1$, of a particular case ($m=0$, $n=1$) of Corollary 4.2 of \cite{Rains2009}.

Three asymptotic results for the gamma functions \eqref{legf2}, and \eqref{lhgf}, will be used, which follow directly from Corollary 3.1, Proposition 2.10, and Corollary 2.3, respectively of \cite{Rains2009}.  First, let $\{x\}\in [0,1)$ denote the fractional part of the real number $x$.  Then in the limit \eqref{hyplim} the lens elliptic gamma function \eqref{legf2} satisfies,
\beq
\label{asymp1}
\ds\frac{\Gamma_e(\epsilon z_1+x,m_1;\epsilon\omega_1,\epsilon\omega_2)\,\Gamma_e(\epsilon z_2-x,m_2;\epsilon\omega_1,\epsilon\omega_2)}{\exp\left\{2\pi\ii\left(R_2(\epsilon z_1,0;\epsilon\omega_1,\epsilon\omega_2)+R_2(\epsilon z_2,0;\epsilon\omega_1,\epsilon\omega_2)\right)\right\}}=O\!\left(\EXP^{2\pi\ii\,\{x\}\{-x\}(z_1+z_2-(\omega_1+\omega_2))/(2\epsilon r\omega_1\omega_2)}\right),
\eeq
where $m_1,m_2\in\mathbb{Z}$, $z_1,z_2\in\mathbb{C}$, $x\in\mathbb{R}$.  Importantly the factor in the denominator on the left hand side is independent of the value of the integers $m_1,m_2$.

Next define
\beq
 G_e(z,m;\omega_1,\omega_2)=\phi_e(z,\llbracket m\rrbracket;\omega_1,\omega_2)+G_{e,1}(z+\llbracket m\rrbracket\omega_1;r\omega_1,\omega_1+\omega_2)+G_{e,1}(z+(r-\llbracket m\rrbracket)\omega_2;r\omega_2,\omega_1+\omega_2)\,,
\eeq
\beq
G_h(z,m;\omega_1,\omega_2)=\phi_h(\llbracket m\rrbracket)+G_{h,1}(z+\llbracket m\rrbracket\omega_1;r\omega_1,\omega_1+\omega_2)+G_{h,1}(z+(r-\llbracket m\rrbracket)\omega_2;r\omega_2,\omega_1+\omega_2)\,,
\eeq
so that
\beq
\Gamma_e(z,m;\omega_1,\omega_2)=\EXP^{G_e(z,m;\omega_1,\omega_2)}\,,\quad\Gamma_h(z,m;\omega_1,\omega_2)=\EXP^{G_h(z,m;\omega_1,\omega_2)}\,.
\eeq
Through the functional relations satisfied by the gamma functions $\Gamma_{e,1}$, and $\Gamma_{h,1}$, the functions $G_e$, and $G_h$, have unique analytic continuations to the complex plane with branch cuts located at the poles and zeroes of the respective functions $\Gamma_e$, and $\Gamma_h$ \cite{Rains2009}.  In the limit \eqref{hyplim}, the function $\Gamma_e$ \eqref{legf2} is related to the function $\Gamma_h$ \eqref{lhgf} by
\beq
\label{asymp2}
-2\pi\ii R_2(z,0;\epsilon\omega_1,\epsilon\omega_2)+G_e(z,m;\epsilon\omega_1,\epsilon\omega_2)-G_h(z,m;\epsilon\omega_1,\epsilon\omega_2)=O\!\left(\EXP^{-2\pi \alpha/\epsilon}\right),
\eeq
where the real number $\alpha$ satisfies
\beq
0<\alpha<\min_{\omega'\in\Omega}\left(\im\left(\frac{-1}{\omega'}\right)-\left|\im\left(\frac{z}{\omega'}\right)\right|\right),
\eeq
for $\Omega=\{r\omega_1,r\omega_2,\omega_1+\omega_2\}$.  The limit of the lens elliptic gamma function \eqref{hyplim2} follows from \eqref{asymp2} for $\epsilon\rightarrow0$.

The asymptotics of $\Gamma_h$ for $z\rightarrow\infty$ are given by
\beq
\label{asymp3}
\begin{array}{c}
\ds-\pi\ii P(z,m;\omega_1,\omega_2)-\phi_h(\llbracket m\rrbracket)+G_h(z,m;\omega_1,\omega_2)=O\!\left(\EXP^{-2\pi\alpha |z|}\right),\\[0.3cm]
\ds+\pi\ii P(-z,m;\omega_1,\omega_2)-\phi_h(\llbracket m\rrbracket)+G_h(-z,m;\omega_1,\omega_2)=O\!\left(\EXP^{-2\pi\alpha |z|}\right),
\end{array}
\eeq
where the real number $\alpha$ satisfies
\beq
0<\alpha<\min_{\omega'\in\Omega}\left(\im\left(-\frac{\EXP^{\ii\arg z}}{\omega'}\right)\right),
\eeq
for $\Omega=\{r\omega_1,r\omega_2,\omega_1+\omega_2\}$, and $P(z,m;\omega_1,\omega_2)$ is defined in terms of $B_\varphi(z,m;\omega_1,\omega_2)$ \eqref{bphidef} as
\beq
2P(z,m;\omega_1,\omega_2)=B_\varphi(-z+(\omega_1+\omega_2)/2;-\ii\omega_1,-\ii\omega_2)\,.
\eeq

The following expansion is also needed for the factor $\lambda$ \eqref{lambdadef}
\beq
\label{asymp4}
1-r\epsilon\sqrt{-\omega_1\omega_2}\,\EXP^{2\pi\ii R_2(0,0,\epsilon\omega_1,\epsilon\omega_2)}(\EXP^{2\pi\ii\omega_1\epsilon};\EXP^{2\pi\ii\omega_1\epsilon})_\infty(\EXP^{2\pi\ii\omega_2\epsilon};\EXP^{2\pi\ii\omega_2\epsilon})_\infty=O(\EXP^{-2\pi\alpha/(\epsilon r)})\,,
\eeq
where the real number $\alpha$ satisfies
\beq
0<\alpha<\min\left(\im\left(\frac{-1}{\omega_1}\right),\im\left(\frac{-1}{\omega_2}\right)\right).
\eeq
From the equations \eqref{asymp1}, \eqref{asymp2}, \eqref{asymp3}, \eqref{asymp4}, the hyperbolic limit from \eqref{eident} to \eqref{hident} follows an identical argument of Theorem 4.1 of \cite{Rains2009}.

Specifically, define $\rho_e(z,y,t_i,u_i;\omega_1,\omega_2)$ to be the integrand of \eqref{eident}
\beq
\rho_e(z,y,t_i,u_i;\omega_1,\omega_2)=\frac{\prod_{i=1}^6\Gamma_e(t_i\pm z,u_i\pm y;\omega_1,\omega_2)}{\Gamma_e(\pm 2z,\pm 2y;\omega_1,\omega_2)}\,.
\eeq
In the limit \eqref{hyplim}, Equation \eqref{asymp1} gives
\beq
\label{limdef}
\begin{array}{r}
\ds\EXP^{4\pi\ii\left(R_2(0,0;\epsilon\omega_1,\epsilon\omega_2)-\sum_{i=1}^6R_2(\epsilon t_i,0;\epsilon\omega_1,\epsilon\omega_2)\right)}\rho_e(z,y,\epsilon t_i,u_i;\epsilon\omega_1,\epsilon\omega_2)\\[0.3cm]
=O\!\left(\EXP^{-2\pi\ii\left(2\{z\}\{-z\}-\{2z\}\{-2z\}/2\right)(\omega_1+\omega_2)/(\epsilon r\omega_1\omega_2)}\right).
\end{array}
\eeq
Since $\im(1/\omega_1+1/\omega_2)<0$, the left hand side of \eqref{limdef} is maximised at the minimum of $2\{z\}\{-z\}-\{2z\}\{-2z\}/2$, which by Lemma 3.3 of \cite{Rains2009}, is at $z=0$.  This allows the restriction of the integrand to the smaller interval $[-1/4,1/4]$ with the introduction of an exponentially small error.  The Equation \eqref{asymp2} may now be used to replace all instances of the lens elliptic gamma functions \eqref{legf2} on the left hand side of \eqref{limdef}, with its hyperbolic analogue \eqref{lhgf}.  The remaining contribution from the $R_2$ polynomials coming from the normalisation functions $\phi_e$ \eqref{ellnorm}, cancels with the contribution from the $R_2$ polynomials appearing on the left hand side of \eqref{limdef}.

Following a change of variables $z\rightarrow \epsilon z$, the use of \eqref{asymp3} reveals that the resulting integrand from \eqref{limdef}
\beq
\rho_h(z,y,t_i,u_i;\omega_1,\omega_2)=\frac{\prod_{i=1}^6\Gamma_h(t_i\pm z,u_i\pm y;\omega_1,\omega_2)}{\Gamma_h(\pm 2z,\pm 2y;\omega_1,\omega_2)}\,,
\eeq
decays exponentially for $z\rightarrow\pm\infty$ as
\beq
\rho_h(z,y,t_i,u_i;\omega_1,\omega_2)=O\!\left(\EXP^{-2\pi\ii\,\left| z\right|\,(\omega_1+\omega_2)/(r\omega_1\omega_2)}\right).
\eeq
This means that the integration of $\rho_h$ may be extended to the entire real line with only exponentially small error, and the final result for the limit $\epsilon\rightarrow0$ of \eqref{limdef} may be written as
\beq
\begin{array}{r}
\ds\lim_{\epsilon\rightarrow0}\EXP^{2\pi\ii\left(3 R_2(0,0,\epsilon\omega_1,\epsilon\omega_2)-2\sum_{i=1}^6R_2(\epsilon t_i,0;\epsilon\omega_1,\epsilon\omega_2)\right)}\lambda\sum_{y=0}^{\floor{r/2}}\varepsilon(y)\int^{\frac{1}{2}}_{-\frac{1}{2}}\!\! dz \,\rho_e(z,y,\epsilon t_i,u_i;\epsilon\omega_1,\epsilon\omega_2)\\[0.6cm]
\ds=\frac{1}{2r\sqrt{-\omega_1\omega_2}}\,\sum_{y=0}^{\floor{r/2}}\varepsilon(y)\int^{\infty}_{-\infty}dz\,\rho_h(z,y,t_i,u_i;\omega_1,\omega_2)\,,
\end{array}
\eeq
with exponentially fast convergence to the right hand side.  As a consequence, the identity \eqref{hident}, follows from the identity \eqref{eident} in the hyperbolic limit \eqref{hyplim} (the limit of the right hand side of \eqref{eident} follows from the use of \eqref{asymp2}), which is what was to be shown.

The star-triangle relation \eqref{hstr} is recovered from \eqref{hident} after setting
\beq
\begin{array}{c}
\ds t_1=+x_1+\ii\alpha\,,\quad t_3=+x_3+\ii\beta\,,\quad t_5=+x_2+\omega_1+\omega_2-\ii(\alpha+\beta)\,,\\[0.3cm]
\ds t_2=-x_1+\ii\alpha\,,\quad t_4=-x_3+\ii\beta\,,\quad t_6=-x_2+\omega_1+\omega_2-\ii(\alpha+\beta)\,,
\end{array}
\eeq
and
\beq
\begin{array}{c}
\ds u_1=+m_1\,,\quad u_3=+m_3\,,\quad u_5=+m_2\,,\\[0.3cm]
\ds u_2=-m_1\,,\quad u_4=-m_3\,,\quad u_6=-m_2\,,
\end{array}
\eeq
and finally substituting $\omega_1\rightarrow\ii\omega_1$, $\omega_2\rightarrow\ii\omega_2$, for $\re(\omega_1),\re(\omega_2)>0$.

The gauge theory equation \eqref{pfidentity} is recovered, up to the different respective normalisations used in \eqref{signfactor} and \eqref{hypnorm}, after the simpler substitution
\beq
t_i=x_i\,,\quad u_i=m_i\,,
\eeq
followed by $\omega_1\rightarrow\ii\omega_1$, $\omega_2\rightarrow\ii\omega_2$, for $\re(\omega_1),\re(\omega_2)>0$.  Some details are given in Appendix \ref{app:modr} on the validity of the identity \eqref{hident} (resp. \eqref{eident}) under such different normalisations of the hyperbolic gamma function \eqref{lhgf} (resp \eqref{legf2}).

\section{The \texorpdfstring{$\!\!\!\!\mod r$} ~~dependence}\label{app:modr}

In this section it will be shown that the dependence on $\llbracket m\rrbracket=m\!\mod r$ in the identities \eqref{eident}, and \eqref{hident}, may be replaced with a dependence on the integer $m$ itself.  This directly follows from the fact that the gamma functions \eqref{legf2}, and \eqref{lhgf}, are unchanged after replacing $\llbracket m\rrbracket$ with $m$.  It is straightforward to show this, through the use of well-known relations for the regular elliptic gamma function \eqref{egf2}, and the associated theta function in the elliptic case, and analogous relations for the hyperbolic gamma function \eqref{hgf} in the hyperbolic case.  This section was motivated by the different form of the identity \eqref{eident}, recently appearing in \cite{rarified}.  The considerations of this section make clear that the univariate case of the elliptic sum/integral identity of \cite{rarified} (Theorem 2 of \cite{rarified}), is equivalent to the identity \eqref{eident} originally obtained by the second author \cite{Kels:2015bda}.

Specifically, redefine the lens elliptic gamma function as
\beq
\label{legf3}
\Gamma_{e}(z,m;\sigma,\tau)=\EXP^{\phi_e(z,m;\sigma,\tau)}\gamma_e(z,m;\sigma,\tau)\,,
\eeq
where $\gamma_e(z,m;\sigma,\tau)$ is given in \eqref{littlegamma}, and $\phi_e(z,m;\sigma,\tau)$ is given in \eqref{ellnorm}, in terms of the function $R_2(z,m;\sigma,\tau)$ now defined as
\beq
R_2(z,m;\sigma,\tau)=R(z+m\sigma;\hat{r}\sigma,\sigma+\tau)+R(z+(\hat{r}-m)\tau;\hat{r}\tau,\sigma+\tau)\,.
\eeq
The $\hat{r}$ is an independent non-zero integer parameter, where $\hat{r}=r$ in \eqref{r2def}.  It will be shown that the identity \eqref{eident} is also satisfied by \eqref{legf3}, and is in fact independent of the value of $\hat{r}$.  The case $\hat{r}=r$ corresponds to the original normalisation of the lens elliptic gamma function used in \cite{Kels:2015bda}, and the case $\hat{r}=1$ corresponds to the recent normalisation appearing in \cite{rarified}.

For now, fix $\hat{r}=r$, and introduce the theta function
\beq
\label{thtdef}
\theta(z;\sigma)=(\EXP^{2\pi\ii z};\p)_\infty\,(\EXP^{-2\pi\ii z}\p;\p)_\infty\,,
\eeq
where $z\in\mathbb{C}$, and $\p$ is defined in \eqref{nomedef}.  The theta function satisfies the identity
\beq
\label{thtident}
\theta(z+k\sigma;\sigma)=\frac{\theta(z;\sigma)}{(-\EXP^{2\pi\ii z})^k\EXP^{2\pi\ii\sigma k(k-1)/2}}\,,\quad k\in\mathbb{Z}\,.
\eeq
The regular elliptic gamma function \eqref{egf2} satisfies the following relations with the theta function \eqref{thtdef}
\beq
\label{egfident}
\begin{array}{c}
\ds\Gamma_{e,1}(z+n\sigma;\sigma,\tau)=\Gamma_{e,1}(z;\sigma,\tau)\prod_{j=0}^{n-1}\theta(z+j\sigma;\tau)\,,\\[0.6cm]
\ds\Gamma_{e,1}(z+n\tau;\sigma,\tau)=\Gamma_{e,1}(z;\sigma,\tau)\prod_{j=0}^{n-1}\theta(z+j\tau;\sigma)\,,
\end{array}
\eeq
for integers $n=1,2,\ldots$.

It may be shown that for $\hat{r}=r$,
\beq
\Gamma_e(z,\llbracket m\rrbracket;\sigma,\tau)=\Gamma_e(z,m;\sigma,\tau)\,,
\eeq
for all integers $m$, and thus \eqref{eident} is unchanged after replacing all integers $\llbracket m\rrbracket$, with $m$.  This follows from the following periodicity property of \eqref{legf3} (for $\hat{r}=r$)
\beq
\label{rperiod}
\Gamma_e(z,m+kr;\sigma,\tau)=\Gamma_e(z,m;\sigma,\tau)\,,\qquad k\in\mathbb{Z}\,,
\eeq
which may be shown directly with the use of \eqref{thtident}, and \eqref{egfident}.  For example, consider the shift $m-ar$, for integers $a\geq1$. Then
\beq
\begin{array}{rcl}
\ds\gamma_e(z,m-ar;\sigma,\tau)&=&\ds\Gamma_{e,1}(z+\sigma(m-ar);r\sigma,\sigma+\tau)\,\Gamma_{e,1}(z+\tau(r-m+ar);r\tau,\sigma+\tau)\\[0.3cm]
&=&\ds\gamma_e(z,m;\sigma,\tau)\prod_{j=0}^{a-1}\frac{\theta(z-\tau(m-(j+1)r);\sigma+\tau)}{\theta(z+\tau(m-(j+1)r);\sigma+\tau)}\,.
\end{array}
\eeq
The ratio of theta functions may be exchanged for an exponential term using \eqref{thtident}, and the desired relation
\beq
\Gamma_e(z,m-ar;\sigma,\tau)=\Gamma_e(z,m;\sigma,\tau)\,,
\eeq
follows after inspecting that the following equality is satisfied
\beq
\EXP^{\phi_e(z,m-ar;\sigma,\tau)}\prod_{j=0}^{a-1}\frac{\theta(z-\tau(m-(j+1)r);\sigma+\tau)}{\theta(z+\sigma(m-(j+1)r);\sigma+\tau)}=\EXP^{\phi_e(z,m;\sigma,\tau)}\,.
\eeq
The case of the shift $m+ar$, for $a\geq1$ is very similar.  Thus the identity \eqref{eident} holds with the lens elliptic gamma function \eqref{legf3}, for the case $\hat{r}=r$.  Note that for the normalisation used in \cite{rarified} (corresponding to $\hat{r}=1$), the analogue of \eqref{rperiod} is a more cumbersome quasi-periodicity condition of the lens elliptic gamma function.

Then after combining all $z$ independent terms coming from the normalisation factors $\phi_e(z,m;\sigma,\tau)$ \eqref{ellnorm}, into a single normalisation factor $\beta(t_i,u_i,y)$, the identity \eqref{eident} may be written in terms of $\gamma_e(z,m;\sigma,\tau)$ \eqref{littlegamma} as
\beq
\label{eident3}
\ds\lambda\sum_{y=0}^{\floor{r/2}}\varepsilon(y)\beta(t_i,u_i,y)\int_0^{1}\frac{dz}{\EXP^{4\pi\ii zy}}\,\frac{\prod_{i=1}^6\gamma_e(t_i\pm z,u_i\pm y;\sigma,\tau)}{\gamma_e(\pm 2z,\pm 2y;\sigma,\tau)}\ds=\!\!\!\!\prod_{1\leq i<j\leq6}\!\!\gamma_e(t_i+t_j,u_i+u_j;\sigma,\tau)\,,
\eeq
where
\beq
\label{finnorm}
\beta(t_i,u_i,y)=\exp\left\{2\pi\ii\left(\sum_{i=1}^6t_iu_i-\left(\sigma-\tau\right)\left(y^2-\frac{1}{2}\sum_{i=1}^6u_i^2\right)\right)\right\}.
\eeq
Observe that the normalisation of the lens elliptic gamma function \eqref{legf3} depends on $\hat{r}$ explicitly only, however the combined normalisation $\beta(t_i,u_i,y)$ \eqref{finnorm} in \eqref{eident3}, is independent of $\hat{r}$.  Thus the identity \eqref{eident} remains invariant after replacing the integer parameter $\hat{r}$ (which has been fixed to $\hat{r}=r$ up to now) appearing in the normalisation factor $\phi_e(z,m;\sigma,\tau)$, by any non-zero integer (generally $\hat{r}$ could be any non-zero complex number).  Consequently \eqref{eident} holds with the function \eqref{legf3} for any non-zero choice of $\hat{r}$, which is what was to be shown.

Now setting $\hat{r}=1$ in $\phi_e(z,m;\sigma,\tau)$ exactly results in the normalisation of $\gamma_e(z,m;\sigma,\tau)$ used in \cite{rarified}.  Thus contrary to what was stated at the end of Section 5 of \cite{rarified}, the identity of Theorem 2 of \cite{rarified} is equivalent to the second authors original identity \cite{Kels:2015bda}, corresponding to the normalisation $\hat{r}=r$.  The normalisation corresponding to $\hat{r}=1$ was chosen in \cite{rarified} so that
\beq
\label{legftoegf}
\Gamma_e(z,m;\sigma,\tau)=\Gamma_{e,1}(z;\sigma,\tau)\,,\quad (r=1)\,,
\eeq
is satisfied for all integers $m$, where $\Gamma_{e,1}(z;\sigma,\tau)$ is the regular elliptic gamma function \eqref{egf2}.  This is also obviously satisfied for the normalisation $\hat{r}=r$.  Due to the periodicity relation \eqref{rperiod} satisfied by the lens elliptic gamma function \eqref{legf3} for $\hat{r}=r$, it is likely that the hypergeometric equations constructed from \eqref{Vfunc} will take a simpler form than in \cite{rarified}, however details of this calculation are beyond the scope of this paper.

The removal of the dependence on $\!\!\!\!\mod r$ is similar at the hyperbolic level.  First redefine the function $\Gamma_h$ by dropping the $\!\!\!\!\mod r$ dependence in \eqref{lhgf}
\beq
\label{lhgf2}
\Gamma_h(z,m;\omega_1,\omega_2)=\EXP^{\phi_h(m)}\gamma_h(z,m;\omega_1,\omega_2)\,,
\eeq
where $\phi_h(m)$, and $\gamma_h(z,m;\omega_1,\omega_2)$ are defined in \eqref{hypnorm}, and \eqref{littlegammah} respectively.

The periodicity property
\beq
\label{hrperiod}
\Gamma_h(z,m+kr;\sigma,\tau)=\Gamma_h(z,m;\sigma,\tau)\,,\qquad k\in\mathbb{Z}\,,
\eeq
follows simply from the following functional relations satisfied by $\Gamma_{h,1}(z;\omega_1,\omega_2)$ \eqref{hgf} 
\beq
\begin{array}{c}
\ds\Gamma_{h,1}(z+n\omega_1;\omega_1,\omega_2)=\Gamma_{h,1}(z;\omega_1,\omega_2)\prod_{j=0}^{n-1}2\sin\frac{\pi(z+j\omega_1)}{\omega_2}\,,\\[0.6cm]
\ds\Gamma_{h,1}(z+n\omega_2;\omega_1,\omega_2)=\Gamma_{h,1}(z;\omega_1,\omega_2)\prod_{j=0}^{n-1}2\sin\frac{\pi(z+j\omega_2)}{\omega_1}\,,
\end{array}
\eeq
for $n=1,2,\ldots$.  Consequently the hyperbolic identity \eqref{hident} is also satisfied by the function \eqref{lhgf2}.  While the normalisation factor \eqref{hypnorm} is required for the periodicity \eqref{hrperiod} of $\Gamma_h$ \eqref{lhgf2}, it may be omitted from \eqref{lhgf2} without affecting the validity of the identity \eqref{hident}.  This follows from the absence of any terms in the exponent of $\beta(t_i,u_i,y)$ \eqref{finnorm} that depend only on the integers $u_i,y$.  As a consequence, the ``sign factor'' normalisation \eqref{signfactor} also cannot make any overall contribution to the identity \eqref{pfidentity} (equivalently \eqref{hident}), and could be dropped from \eqref{sbdef} without affecting the validity of the latter identity.

\end{appendices}

\bibliographystyle{utphys}
\bibliography{star}

\end{document}